\documentstyle[aps,psfig]{revtex}

\begin{document}

\title{{\bf Phase-Locked Spatial Domains and Bloch Domain Walls \\
in Type-II
Optical Parametric Oscillators}}

\author{{\it Gonzalo Iz\'us \cite{gonzalo} and Maxi San Miguel}}
\address{Instituto Mediterr\'aneo de Estudios Avanzados, IMEDEA (CSIC-UIB)
\cite{www},\\
E-07071 Palma de Mallorca, Spain}

\author{{\it Marco Santagiustina}}
\address{Istituto Nazionale di Fisica della Materia,
Dipartimento di Elettronica e Informatica,\\
Universit\`a di Padova,
via Gradenigo 6/a, 35131 Padova, Italy}

\maketitle

\vskip 1.5cm

\begin{abstract}
We study the role of transverse spatial degrees of freedom in the
dynamics of signal-idler phase locked states in type-II Optical
Parametric Oscillators. Phase locking stems from signal-idler
polarization coupling which arises if the cavity birefringence
and/or dichroism is not matched to the nonlinear crystal
birefringence. Spontaneous Bloch domain wall formation is
theoretically predicted and numerically studied. Bloch walls
connect, by means of a polarization transformation, homogeneous
regions of self-phase locked solutions. The parameter range for
their existence is analytically found. The polarization properties
and the dynamics of walls in one- and two transverse spatial
dimensions is explained. Transition from Bloch to Ising walls is
characterized, the control parameter being the linear coupling
strength. Wall dynamics governs spatiotemporal dynamical states of
the system, which include transient curvature driven domain
growth, persistent dynamics dominated by spiraling defects for
Bloch walls, and labyrinthine pattern formation for Ising walls.
\end{abstract}
\newpage

\section{INTRODUCTION}

Optical Parametric Oscillators (OPO) are versatile nonlinear
optical devices \cite{JOSAB99} with a variety of possible
applications including useful alternatives to lasers and the
generation of light with non classical properties \cite{quantum,Lugiatoreview}.
For optical cavities with large Fresnel number, they have also
become a paradigm for the study of transverse pattern formation
that arise in optical systems as a consequence of diffraction and
nonlinearity \cite{pinos,oppo94,valcarcel96}. Experimental
observations of such patterns have been reported \cite{fabr99}.
Recent interest in these transverse structures in OPOs arise from
the study of macroscopic manifestations of quantum phenomena in
the spatial correlations present in these patterns
\cite{Lugiatoreview}, as well as from the study of spatially
localized structures, such as domain walls and cavity solitons
\cite{oppo01,Berre,op2,tril97,long97,op1,stal98,oppo99}, with
possible applications in all-optical signal processing.

In type-I OPO the signal and idler fields generated in the down
conversion process have the same state of linear polarization. In
type-II OPO signal and idler are orthogonally polarized. The
additional vectorial degree of freedom of type-II OPO is very
interesting from the point of view of possible new nonlinear
phenomena. An interesting example of these new possibilities has
been recently observed experimentally and described theoretically
\cite{Mason98,Fabre00} when considering a direct intra-cavity
polarization coupling: It is possible to reach a situation of
frequency degeneracy and phase locking between between the
orthogonally polarized signal and idler fields. This is important
because, without the direct polarization coupling, a type-II OPO
remains non degenerate at frequency degeneracy because of
polarization. In this phase locked situation the polarization of
the output field is determined by the locked value of the relative
phase between signal and idler, which can be tuned by changes of
experimentally accessible parameters. This device has been
proposed as a candidate to generate  bright quantum entangled
states. Our general aim in this paper is to consider such phase
locked states in a cavity of large Fresnel number, and to explore
how the transverse spatial degrees freedom enter in the
description of the phenomenon. We find that equivalent phase
locked solutions grow locally, forming spatial domains separated
by domain walls. We study the nature and dynamics of these domain
walls.

When considering transverse spatial degrees of freedom in a
type-II OPO without direct polarization coupling there are two
different regimes. In one of them, characterized by an effective
negative detuning \cite{Nos99}, a finite wave number is selected
at threshold. In a second regime to be considered in this paper,
and which occurs for the opposite sign of detuning, homogeneous
solutions are selected at threshold. However, there is a
continuous of possible solutions with arbitrary relative phase
between signal and idler. Therefore, there are no possible domain
walls. This is different of what happens in type-I OPO in which,
for the equivalent regime of detunings,  homogeneous solutions
with two possible opposite phases can be selected. As a
consequence, spatial phase domains appear in the system separated
by domain walls
\cite{oppo01,Berre,tril97,long97,op1,stal98,oppo99,kutz99,oppo98}.
Such domain walls are of Ising type, that is fronts for which the
field vanishes at the core of the wall \cite{Coullet90}. Domain
walls with the same symmetry properties have been also reported
for a variety of other optical systems
\cite{thorsten,lede98,mand98,stal98b,rafa,tara98,roza96}. Direct
polarization coupling in type-II OPO breaks the invariance under
changes of relative phase, allowing for the formation of domain
walls. The novelty here is that these walls can be either of Ising
or Bloch type \cite{op2,Coullet90}. Differences between Ising and
Bloch walls are that there are two equivalent Bloch ways to
connect two spatial domains (symmetry breaking) and that Bloch
walls can move spontaneously, leading to complicated persistent
dynamical states of the system. The transition from Ising to Bloch
walls is controlled by the strength of the polarization coupling.
Bloch walls have been recently predicted in other optical systems
\cite{bloch}.

Direct polarization coupling between signal and idler in type-II
OPO  has been discussed in the literature
\cite{Mason98,Fabre00,kimb90,Lee90} by considering the insertion
in the optical cavity of wave-plates (such as quarter-wave or
half-wave). In these previous studies the transverse spatial
degrees of freedom were not considered. The generic phenomena
described in this paper are expected for any form of direct
polarization coupling. However, we address specifically the
signal-idler coupling arising from birefringence and dichroism of
the cavity mirrors, although our general equations give a general
representation of possible forms of polarization coupling. A small
amount of birefringence or dichroism is always present due to weak
cavity imperfections and therefore the phenomenon considered here
should be generally present in type-II OPO's.

The paper is organized as follows. Section 2 presents our general
model equations, which are derived in detail in an Appendix. In
section 3 we calculate the OPO threshold, we describe the possible
stationary phase-locked homogeneous solutions, and their
polarization properties are characterized in terms of the Stokes
parameters. In Section 4 we discuss domain walls in one transverse
spatial dimension (1D): Ising and Bloch walls, their dynamics and
the Bloch-Ising transition are characterized. We also describe the
polarization properties of these walls. Sections 5 and 6 describe
the dynamical states in 2 transverse spatial dimensions (2D)
governed, respectively, by Bloch or Ising walls. Our main
conclusions are summarized in Section 7.

\section{EQUATIONS FOR A TYPE-II OPO WITH DIRECT POLARIZATION COUPLING}

A type-II OPO that consists of a ring optical resonator, filled
with a birefringent, nonlinear quadratic medium will be
considered. The device is externally pumped by a laser beam,
uniform in the plane transverse to the cavity longitudinal axes
and of frequency $\omega_p$. We take into account effects of
birefringence and dichroism, that can be due either to small
imperfections of the cavity mirrors or to weakly birefringent (e.g. wave
plates) or dichroic optical devices inserted in the optical
cavity. The derivation of the governing equations is presented,
for clarity, in the appendix. For the sake of simplicity, the
cavity birefringence and dichroism are supposed to be only due to
one of the resonator mirrors. Note also that, in general, the
mirror principal axes (i.e. those along which the Jones matrix
\cite{phot} that represents the polarization transformation is
diagonal) are rotated with respect to the principal axes of the
crystal (i.e. those along which the susceptibility matrix is
diagonal). This rotation angle ($\phi$) is an important
experimental parameter through which the strength of the effects
we describe below can be controlled. We note that the equations
are obtained in the mean field, paraxial and single longitudinal
mode approximation for all the fields involved.
\newline
The equations that describe the time evolution couple together
four field envelopes that depend on the transverse coordinates
$x,y$: the linear polarization components of the intra cavity
field at the pump frequency: $B_{x,y}(x,y,t)$; the signal and
idler fields: $A_{x,y}(x,y,t)$. The signal and the idler can be
either frequency degenerate or non-degenerate, depending on the
frequency selection rules imposed by the combined effects of the
parametric down-conversion, the cavity resonances and the
phase-matching \cite{falk71,byer91,fabr93}, but they are always
polarization non-degenerate (type-II interaction). Hereafter the
frequency degenerate (or quasi-degenerate) case, that is routinely
obtained by tuning the phase-matching conditions \cite{opos}, will
be considered. Moreover, with no loss of generality we set $A_x$
and $B_x$ to be ordinary polarized beams and $A_y,B_y$ to be
extraordinary polarized \cite{Nos99}. Then the equations
describing the OPO are:
\begin{eqnarray}
\partial_t B_x &=& \gamma_x' [-(1+i \Delta_x')
B_x + i \alpha_x' \nabla^2 B_x +c'_x B_y + 2i K_0 A_x A_y + E_0] \nonumber \\
\partial_t B_y &=& \gamma_y' [-(1+i \Delta_y') B_y + i \alpha_y' \nabla^2 B_y +
c'_y B_x] \nonumber \\
\partial_t A_x &=& \gamma_x [-(1+i \Delta_x) A_x + i \alpha_x \nabla^2 A_x +
c_x A_y + i K_0 A_y^* B_x ] \nonumber \\
\partial_t A_y &=& \gamma_y [-(1+i \Delta_y) A_y + i \alpha_y \nabla^2 A_y +
c_y A_x + i K_0 A_x^* B_x]
\label{master}
\end{eqnarray}
The coefficients $\gamma_{x,y}, \gamma_{x,y}'$ (effective cavity
decay rates), $\Delta_{x,y},\Delta_{x,y}'$ (effective cavity mode
detunings) and $\alpha_{x,y},\alpha_{x,y}'$ (diffraction
coefficients) are defined in the appendix (see eqs.
(\ref{coeffs})). Some general remarks are worth to be made to show
differences and common features between these coefficients and
those previously defined for a "perfect" cavity (see for example
\cite{oppo94,Nos99}). Due to the birefringence of the nonlinear
crystal and the dichroism of the cavity, the coefficients of
equivalent terms, in different equations, are slightly different,
even for frequency degenerate fields. In fact they all depend on
the relative refractive index and mirror transmittivity that are
polarization dependent (see eqs. (\ref{coeffs}) for details). The
cavity birefringence can also cause the effective detuning
coefficients, $\Delta_{x}$, $\Delta_{y}$ ($\Delta_{x}'$ , $\Delta_{y}'$), 
to be different, even
at frequency degeneracy. Other parameters are the nonlinearity
$K_0$ (defined by eq. (\ref{nlcoeff})) and the injected pump $E_0$
that is taken as a real number. This gives no loss of generality
because it is equivalent to fix a common phase reference for all
fields. For the sake of simplicity, we take the pump to be
linearly polarized in a direction parallel to the phase-matched
component of the intra cavity field $B_x$. Hence, the highly
mismatch component $B_y$ neither is pumped nor is nonlinearly
coupled with other components. It is therefore very weakly
involved in the dynamics, in spite of the linear coupling with
$B_x$.
\newline
The linear coupling coefficients $c_{x,y}, c'_{x,y}$ account for
the dichroism and the birefringence of the cavity. They read:
\begin{equation}
c_{x,y}=\frac{p+i\delta}{T \pm p \, cos(2 \phi)} sin(2 \phi) , \;
c'_{x,y}=\frac{p'+i\delta'}{T' \pm p' cos(2 \phi)} sin(2\phi)
\label{coupling}
\end{equation}
where the plus (minus) sign applies for the $x$ ($y$) polarized
component. Although their derivation and the exact relation with
the physical parameters describing the cavity can be found in the
appendix it is useful to describe them briefly at this stage. The mirror
dichroism is represented by the ratio between the difference of
the reflectivity and the average reflectivity at a certain
frequency: respectively $2p$ for $A_{x,y}$ (signal and idler) and
$2p'$ for $B_{x,y}$ (pump components). The mirror birefringence
causes a different phase change: $2\delta$ for $A_{x,y}$ and
$2\delta'$ for $B_{x,y}$. Finally $T$ ($T'$) is the average
transmittivity of the signal/idler (pump components) and $\phi$ is
the relative angle of rotation between the crystal and cavity
birefringence axes (the axes of dichroism are supposed to coincide
with those of cavity birefringence for the sake of simplicity). We
note that similar linear coupling terms between signal and idler,
were previously considered \cite{Mason98,Fabre00} by considering the
insertion of wave-plates (such as quarter-wave or half-wave) in
the cavity of a type-II OPO. In these cases the general relation
$c_x=-c_y^*$ is satisfied.

\section{Threshold analysis and homogeneous phase locked solutions}

\subsection{Threshold analysis}

The trivial solution of eqs. (\ref{master}) corresponds to the
case in which a type-II OPO is below the threshold of signal
generation. It is given by:
\begin{eqnarray}
A_{x} & = & A_{y}= 0  \nonumber \\
B_x & =& \tilde{c} E_0 \nonumber \\
B_y & = & \frac{c_y' \tilde{c} E_0}{1+i\Delta_y'}
\label{trivial}
\end{eqnarray}
where $\tilde{c}=c^r+ic^i=(1+i\Delta_y')/[1-\Delta_x' \Delta_y'-
c_x' c_y'+i(\Delta_x'+\Delta_y')]$. The threshold for instability
is determined by linearizing eqs.(\ref{master}) around this
solution and looking for values of the bifurcation parameter $E_0$
(the pump amplitude) for which perturbations grow. The general
type of perturbation is given, as usual, by plane waves
$\exp(i{\vec q} \cdot {\vec r}-\lambda t)$, where $\lambda({\vec
q})$ is the growth rate of the perturbation and ${\vec q}$ is its
transverse wave vector. We first recall the main results of the
analysis for $c_{x,y}=c_{x,y}'=0$ \cite{Nos99}. In that case the
trivial solution is stable for $E_0<E_c$, where:
\begin{equation}
E_c=\frac{(1+i \Delta_x')}{K_0} \, \sqrt{1+\widetilde{\Delta}^2}
\end{equation}
and where the effective detuning $\widetilde{\Delta}$ is defined
as:
\begin{equation}
\widetilde{\Delta}= \frac{\gamma_x
\Delta_x+ \gamma_y\Delta_y}{\gamma_x+\gamma_y}
\end{equation}
If the pump amplitude exceeds $E_c$, the steady state becomes
unstable and the signal and idler fields are generated. In
particular, for negative effective detuning, pattern formation
occurs, as studied in ref. \cite{Nos99}: for this reason hereafter
only the case $\widetilde{\Delta}>0$ will be considered. In this case
there is a Hopf bifurcation in which homogeneous perturbations
${\vec q}=0$ with
\begin{equation}
Im(\lambda)=\omega= - \frac{\gamma_x \gamma_y}{\gamma_x+\gamma_y}
(\Delta_x-\Delta_y)
\label{frequency}
\end{equation}
have the largest growth rate. At threshold, a family of
homogeneous oscillating solutions bifurcate from the trivial
steady-state. For $c_{x,y}=0$, eqs. (\ref{master}) are invariant
under the transformation $A_x\rightarrow A_x \exp{(i\phi)},
A_y\rightarrow A_y \exp{(-i\phi)}$, and therefore the relative phase
between the homogeneous oscillating solutions $ A_x, A_y$ is arbitrary.

The introduction of the polarization coupling  $c_{x,y} \neq 0$
breaks the invariance under changes of the relative phase. One
expects that such coupling should be able to produce phase-locked
homogeneous stationary (i.e., zero frequency) solutions above
threshold. The linear stability analysis of (\ref{trivial}) is
rather cumbersome when $c_{x,y} \neq 0$, and simple analytical
expressions for the threshold analysis are not found. It is in
this case more convenient to determine a threshold through the
condition of existence of the relative phase locked solutions. We
consider the special case $c_x=c_y=c_x^r+ic_x^i$ ($(c_x^r,c_x^i)
\in {\cal R}$) which through eqs. (\ref{coupling}) can be seen to
correspond either to set $\phi=\pm \pi/4$ or $\phi=\pm 3 \pi/4$
(i.e. to fix the angle to the value that maximizes the coupling
strength) or to set $p=c_x^r=0$ (i.e. no dichroism). We note that
setting $p=0$ reproduces a particular case of the
polarization coupling considered in \cite{Mason98,Fabre00}.

Let us call $\bar A_x=a_x \, exp(i \psi_x)$ and $\bar A_y=a_y \, exp(i
\psi_y)$ the homogeneous stationary solutions; by substituting
such formulas into eqs. \ref{master} one gets:
\begin{eqnarray}
sin(\psi_y-\psi_x)&=&\frac{\Delta_x-\Delta_y}{2
\sqrt{(c_x^r+c_x^i \Delta_x)(c_x^r+c_x^i \Delta_y)}} \nonumber \\
cos(\psi_y+\psi_x)&=&\frac{1}{2K_0 E_0 |\tilde{c}|^2 \sqrt{\Gamma}} [(\Delta_x +
\Gamma \Delta_y) c^r -(1+\Gamma) c^i - 2 \sqrt{\Gamma} cos (\psi_y-\psi_x) (c_x^i c^r +
c_x^r c^i)] \nonumber \\
a_x^2&=&\frac{1}{4 K_0^2 c^r \Gamma} [ 2 K_0 E_0 \sqrt{\Gamma}
(c^r sin (\psi_y+\psi_x) - c^i cos(\psi_y+\psi_x) ) + 2 \sqrt{\Gamma}
cos(\psi_y-\psi_x) -(1+\Gamma)] \nonumber \\
a_y^2 &=& \Gamma a_x^2
\label{solutions}
\end{eqnarray}
where $\Gamma=(c_x^r+c_x^i \Delta_x)/(c_x^r+c_x^i \Delta_y)$.

>From eqs. (\ref{solutions}) we find two conditions for the
existence of the phase-locked homogeneous stationary solutions.
The first one comes from the fact that the phase difference among
the solutions (see the first of eqs. (\ref{solutions})) is real only
if the modulus of the right hand side is less than one, that is if
\begin{equation}
(\Delta_y-\Delta_x)^2 \le 4 (c_x^r+c_x^i \Delta_x)(c_x^r+c_x^i \Delta_y)
\label{condition}
\end{equation}
This boundary in the complex plane $c_x^r,c_x^i$ defines the limit
of the locking regime. Inside the boundary stationary solutions do
not exist. Physically speaking, the locking condition means that
the stationary phase-locked solutions exist whenever the direct
polarization coupling breaking is large compared with
the difference in detunings. When the condition (\ref{condition})
is not satisfied, numerical solutions show that there are still
homogeneous states but their phase varies periodically with time.
Such solutions indicate the persistence of the Hopf bifurcation
found for $c_{x,y} = 0$ when the polarization coupling is small.

The second condition for the existence of solutions refers to the
pump value above which there is signal and idler generation. The
threshold $E_c$ can be determined by setting $a_x=0$ into eqs.
(\ref{solutions}) and solving for $E_0$. The final result is:
\begin{equation}
E_c^2=\frac{1}{4 K_0^2 \Gamma |\tilde{c}|^2} \{ (1+\Gamma)^2 +
(\Delta_x + \Gamma \Delta_y)^2 + 4 \Gamma |c_x|^2
cos^2(\psi_y-\psi_x) - 4 \sqrt{\Gamma} cos(\psi_y-\psi_x)
[c_x^r(1+\Gamma) + c_x^i (\Delta_x + \Gamma \Delta_y) ] \}
\label{thresh}
\end{equation}
The classification of the solutions found above this threshold is
easier to understand considering the case $\Delta_x=\Delta_y$ for
which the condition (\ref{condition}) is automatically satisfied.
In this case and regardless of the value of $c_{x,y}$ the relative
phase shift between phase-locked signal and idler can be either 0
(in-phase solution) or $\pi$ (out-of-phase solution) and the
amplitude of the fields is equal ($a_x=a_y$) since $\Gamma=1$.
Once the phase shift $\psi_y-\psi_x$ is known it can be
substituted into the second of eqs. (\ref{solutions}) that can be
solved for $\psi_y+\psi_x$. In principle two solutions exist for
each phase difference (in- and out-of-phase cases), due to the
fact that the $arcos$ is a multi-valued function in the range
$[-\pi, \pi]$. However, if the negative solutions for
$\psi_y+\psi_x$ are replaced into the third of eqs.
(\ref{solutions}) the result is $a_x^2<0$ $\forall E_0$. Therefore
only positive solutions of the angle sum are to be taken to
guarantee that $a_x^2>0$ above a certain threshold $E_c$. By
substituting the phase difference and sum into the third equation
the amplitude is finally found. The sign of $a_x$ can be either
positive or negative, i.e. there are two equivalent possible
solutions for the in-phase case and two for the out-of-phase case.
The existence of these two equivalent solutions is a consequence
of the symmetry $(A_x,A_y) \rightarrow -(A_x,A_y)$ of eqs.
(\ref{master}) which is preserved for $c_{x,y}\neq 0$. In summary,
we find in-phase solutions ($A_x=A_y$) and out-of-phase solutions
($A_x=-A_y$). Each of the in- and  out-of-phase cases include two
equivalent solutions which we denote by the $+$ and $-$ solutions
satisfying $A_{x,y}^+=-A_{x,y}^-$.

In the general case  with $\Delta_x \neq \Delta_y$ solutions are
no longer strictly in- or out-of-phase. Nonetheless, well within
the phase-locked regime where the detuning coefficients are small
compared with the strength of the polarization coupling, the
solutions one finds are close to being in- or out-of-phase.
Therefore we will still use in this situation the names of in- and
out-of-phase solutions even if this is not generally rigorous.
Other situations can occur close to the limit of the locking
regime fixed by eq. (\ref{condition}). For example, for $c_x=c_y$
and a purely dichroic mirror, $c_x^i=0$ so that $\Gamma=1$ and
signal and idler have the same amplitude $a_x=a_y$. At the onset
of the locking regime $\mid \Delta_x - \Delta_y \mid = 2 \mid
c_x^r \mid$, and it follows from the first of eqs.
(\ref{solutions}) that the two fields are locked at a phase
difference $\psi_y - \psi_x = \pm \pi/2$. In any case, for each
locked value of the phase difference there are two equivalent  $+$
and $-$ solutions satisfying $A_{x,y}^+=-A_{x,y}^-$
\newline
A bifurcation diagram for the homogeneous solutions for a generic
case with $\Delta_x \neq \Delta_y$ is presented in figure
\ref{figbifur}. The selection of the solution that actually
bifurcates, either in-phase or out-of-phase, is determined by the
relative value of the threshold $E_c$ for each solution. Eq.
(\ref{thresh}) shows that the threshold is lower if the term
proportional to $cos(\psi_y-\psi_x)$ is positive; for the in-phase
solution this occurs if $c_x^r>0, c_x^i>0$ and vice versa for the
out-of-phase solution if $c_x^r<0, c_x^i<0$. An example of the
thresholds calculated for the in-phase and out-of phase solutions,
as a function of $c_x^r=c_x^i$, and for the same detuning values
than in figure \ref{figbifur} are shown in figure \ref{figthresh}.
Note that there is range of values of the coupling $c_{x,y}$ in
which eq. (\ref{condition}) is not satisfied and there is no
threshold for the emergence of the phase locked solutions. But in
the range in which (\ref{condition}) holds, the lower threshold
decreases as the coupling strength $|c_{x,y}|$ increases, allowing
parametric down-conversion for lower values of the external pump
with respect to the reference case ($c_{x,y}=0$). Note also that
the roles of the two solutions are exchanged if the sign of
$c_{x,y}$ is reversed. We finally note that our numerical
integrations show that when switching-on the pump to a value for
which both the in-phase and out-of-phase solutions exist, the
solution which is always selected is the one with lowest
threshold, while the other is unstable. Therefore, in practice we
only find the two equivalent solutions of lowest threshold.

\subsection{Polarization properties of the phase locked solutions}

An important question is the polarization state of the
phase-locked homogeneous stationary solutions that we have just
described. It is useful to consider the polarization
representation given by the normalized Stokes parameters, defined
as \cite{phot}:
\begin{eqnarray}
s_1&=&\frac{|A_x|^2-|A_y|^2}{|A_x|^2+|A_y|^2} \nonumber \\
s_2&=&\frac{A_x A_y^*+ A_x^* A_y}{|A_x|^2+|A_y|^2} \nonumber \\
s_3&=&\frac{-i(A_x A_y^*- A_x^* A_y)}{|A_x|^2+|A_y|^2}
\label{stokes}
\end{eqnarray}
These real parameters are sufficient to characterize any state of
polarization of a monochromatic field by assigning to the field a
point in the Poincare sphere. The equator of the sphere ($s_3=0$)
corresponds to linearly polarized states and the poles of the
sphere ($s_1=s_2=0, s_3=\pm 1$) correspond to states of opposite circular
polarization. By replacing the homogeneous solutions in eqs. (\ref{stokes}), 
the state
of polarization of our phase locked solutions is represented by:
\begin{eqnarray}
\bar s_1&=&\frac{1-\Gamma}{1+\Gamma} \nonumber \\ \bar
s_2&=&\frac{2\sqrt{\Gamma}}{1+\Gamma} cos(\psi_y-\psi_x) \nonumber
\\ \bar s_3&=&- \frac{2\sqrt{\Gamma}}{1+\Gamma} sin(\psi_y-\psi_x)
\label{pol-state}
\end{eqnarray}
These equations show that the polarization state of the optical
field is determined by locked value of the phase difference of
signal and idler.  For $\Delta_x=\Delta_y$, $\Gamma=1$ and $\sin
(\psi_y - \psi_x)=0$ so that $(\bar s_1, \bar s_2, \bar
s_3)=(0,\pm 1,0)$, where the plus (minus) sign applies for the in-
(out-of) phase solution. This means that the two possible
phase-locked solutions are actually linear and orthogonal
polarizations whose azimuth angles are $\theta=atan(\bar s_2 /
\bar s_1)/2=\pm \pi/4$. We mentioned before that there are two
equivalent solutions for each the in- and out-of-phase solutions.
These correspond to linearly polarized states along the same
direction, but in opposite senses, and they have the same Stokes
parameters. The Stokes parameters are determined by the relative
phase, while the two equivalent solutions have a different global
phase. For example, in the two equivalent in-phase solutions $
\bar s_2=1$  and $Re (A_x)= Re (A_y)$, but in one of the solutions $Re
(A_x) >0$ and in the other one $Re (A_x) <0$.

When detunings are different, the homogeneous solutions become
elliptically polarized beams ($\bar s_3\neq 0$). However if
detunings are small ($c_x^i \Delta_{x,y}<<c_x^r$) still
$\Gamma\simeq 1$ and well within the locking regime
$sin(\psi_y-\psi_x)\simeq 0$. In these circumstances the state of
the beam is close to be linearly polarized ($\bar s_3 \simeq 0$)
with azimuth angles close to $\pm \pi/4$. However, it is important
to note that changing the detuning parameters and the strength of
the polarization coupling it is possible to explore arbitrary
states of polarization. These states are determined from eq.
(\ref{pol-state}) in terms of the phase difference of the locked
state. For example, in the case mentioned above of a purely
dichroic mirror, $c_x^i=0$, we have $\Gamma=1$ and therefore $\bar
s_1 = 0$. In this case, and at the onset of the locking regime,
$\bar s_2 = 0, \bar s_3 = \pm 1$ so that the locked solution is
circularly polarized. Going into the locking regime the
polarization state will evolve towards linearly polarized states
but keeping $\bar s_1 = 0$.

The threshold decrease due to the polarization coupling discussed
above has now a simple physical interpretation. Let us recall that
the condition $c_x=c_y$ means that the relative angle of
inclination $\phi$ must be one of these values: $[\pm \pi/4, \pm 3
\pi/4]$. In particular, through eqs. (\ref{coupling}), $c_x^r$
($c_x^i$) is positive for $p>0$ ($\delta>0$) and $\phi=\pi/4, -3
\pi/4$ or $p<0$ ($\delta<0$) and $\phi=-\pi/4, 3\pi/4$. When $A_x$
and $A_y$ are in phase, the total field is linearly polarized at
$\pi/4$ (or $\-3 \pi/4$) radiants with respect to the crystal
axes. The conclusion that can be taken is that the intra cavity
field is actually oriented along one of the principal axes of the
cavity birefringence-dichroism. The same occurs for the
out-of-phase solution that is a beam linearly polarized at an
angle $-\pi/4$ (or $3 \pi/4$) radiants and then the role of
coefficients is exchanged. So, the polarization selected is always
aligned with one of the cavity principal axes.

\section{PHASE POLARIZATION DOMAIN WALLS IN ONE TRANSVERSE DIMENSION:
 BLOCH-ISING TRANSITION}

In the previous section we have discussed the existence of two
equivalent homogeneous solutions which we named as + and -
solution. These are the solutions with lowest threshold, while we
mentioned that other solutions of higher threshold are seen to be
unstable. When the OPO switches-on after setting the pump to a
value above its threshold value, given by (\ref{thresh}), either
the + or - solution can be selected since they have the same
growth rate. When taking into account the transverse spatial
degrees of freedom, this selection, or spontaneous symmetry
breaking of the homogeneous solution, can be local, with a
different solution emerging in different spatial regions. It is
then expected to find domain walls that separate the spatial
domains with different but equivalent solutions. For either the
signal or idler what changes when going from one solution to the
other is just a sign. For example $A_x$ takes values $\tilde A$ and
$- \tilde A$ at opposite sides of the wall. Therefore the walls can
be considered as phase walls of a complex field like the ones
described for type-I degenerate OPO \cite{tril97,op1,oppo99}. When
considering signal and idler the domain wall separates two
solutions with polarization properties determined by the locking
of the relative phase of the two fields. One might then talk about
polarization walls. However, we already mentioned that the + and -
solution have the same Stokes parameters, but there is a change in
the global phase of the polarization state. In this sense we refer
to these walls as phase polarization walls. In any case, the
polarization state might present interesting features in the core
of the wall.

Phase domain walls for a complex field can be of Ising or Bloch
type \cite{op2,Coullet90}. As a general characterization, in an
Ising wall there is a single field profile connecting one
homogeneous solution with a second equivalent one, while we talk
of Bloch type wall when there are two different field profiles
(walls) connecting the two solutions. A Bloch wall implies,
therefore, spontaneous symmetry breaking for the domain wall. In
the following we study Ising and Bloch domain walls, the
transition between them, and the polarization properties in one
transverse spatial dimension. The characterization of some
properties of these walls is much more clear in one dimension.
Other features associated with two-dimensional phenomena are
postponed to the following sections.

Numerical integrations \cite{numerical} of eqs. (\ref{master})
confirm that stationary uniform domains of the + or - solutions
form spontaneously starting from a randomly and weakly perturbed
trivial steady-state (\ref{trivial}). Well within the locking
regime, the domain walls are of the Ising type, but changing the
values of $c_{x,y}$ and moving towards the boundary of the locking
regime we find a transition from from Ising to Bloch domain walls
\cite{op2}. An example of an IW, is presented in figure
\ref{figiw-1d}a. It connects the + solution at $x \rightarrow
-\infty$ with the + solution at $x \rightarrow \infty$. By
plotting the numerically obtained solution in the complex plane
($Re(A_x), Im(A_x)$) (figure \ref{figiw-1d}b) it is clear to
observe that the IW is characterized by a zero crossing of the
field. An example of 1D optical BW is instead given in figure
\ref{figbw-1d}, a) and b); note that the field amplitude
(represented by the vector modulus in the complex plane) never
goes to zero and the wall consists of an almost pure phase
rotation of $\pi$ radiants. The phase can rotate in two possible
senses along the interface, clockwise or counterclockwise in the
complex plane. This characteristic is usually called the wall
chirality and it is defined to be positive for clockwise rotation,
 or negative for counterclockwise rotation. Therefore, there exist
two equivalent domain  walls of opposite chirality for $A_x$. One
of the two appears by spontaneous symmetry breaking. In the
example of figure \ref{figbw-1d} the wall of negative chirality
for $A_x$ is selected.

In the example of the Ising wall of Fig. \ref{figiw-1d} parameters
are such that we are well within the locking regime and the domain
wall connects homogeneous equivalent in-phase solutions in which
the locked phase difference is close to zero ($\psi_y - \psi_x \simeq
-0.122$). The shape of the field $A_y$ across the domain wall is
similar to the one of $A_x$. The situation is different for the
example of Bloch wall in Fig. \ref{figbw-1d}. Parameters correspond
here to a situation close to the boundary of phase-locking and the
homogeneous + and - solutions have a locked phase difference
$\psi_y - \psi_x = \pi/2$. In addition, Bloch walls for this
system are characterized by the fact that the wall profile for the
field $A_y$ has always opposite chirality to that of $A_x$, as
seen in the example of Fig. \ref{figbw-1d}b.

The polarization characteristics of Ising and Bloch domain walls
are very different. For an Ising wall the Stokes parameters are
seen to remain constant across the core of the wall. This is due
to the fact that the phase difference $\psi_y - \psi_x$ remains
fixed to its locked value while going from the - to the + solution
across the wall. On the contrary, in a Bloch wall the locked value
of the phase difference $\psi_y - \psi_x$ is a function of the
position while moving form one side to the other of the wall. The
consequence is that the Stokes parameters have a nontrivial space
dependence across the wall determining peculiar polarization
characteristics of the core of the wall. As examples of such
polarization characteristics we show in Fig. \ref{figstokes} the
variation of the Stokes parameters across two examples of Bloch
domain walls. In Fig. \ref{figstokes}a we consider a Bloch wall
which connects two linearly polarized states. In
Fig. \ref{figstokes}b,  which corresponds to the wall of
Fig. \ref{figbw-1d}, the wall connects two elliptically polarized
states which are close to being circularly polarized.
\newline
In Fig. \ref{figstokes}a the two asymptotic states for
$x\rightarrow \pm\infty$ are in-phase solutions characterized by
$(\bar s_1, \bar s_2, \bar s_3)=(0,1,0)$ corresponding to a
linearly polarized state of azimuth $\theta=\pi/4$. At the core of
the wall $(\bar s_1, \bar s_2, \bar s_3)=(0,-1,0)$ which
corresponds to a state of orthogonal linear polarization. Note
that as the phase solution rotates $\pi$ radiants the Stokes
parameters $s_1,s_2, s_3$ return to the initial values, i.e. the
polarization of the total field is the same on each side of the
wall, as previously mentioned. Along the wall the polarization
changes, the field becoming elliptically polarized, but not in an
arbitrary manner. The transformation is forced to occur for $s_1
\simeq 0$, i.e. the azimuth angle of the ellipse is practically
fixed to $\theta = \pm \pi/4$. For $x<0$, and close to the core of
the wall, $s_3 \simeq -1$, indicating a state close to being left
circularly polarized, while for $x>0$, and close to the core of
the wall, $s_3 \simeq 1$, indicating a state close to being right
circularly polarized. In terms of the Poincare sphere, the
representative point moves from a point in the equator through the
vicinity of the south pole to the opposite point in the equator,
and back to the original point through the vicinity of the north
pole. The change of ellipticity across the wall, $\eta(x)=\arcsin
(s_3) / 2$, yields a natural interpretation of the chirality: a BW
of negative chirality, like the one shown here, means that the
ellipticity of the polarization state first decreases to a maximum
negative value going as we move to the other side of the wall to a
maximum positive value. For a BW of positive chirality the
excursion in ellipticity goes in the opposite direction.

In Fig. \ref{figstokes}b the variation of the Stokes parameters
indicates a sequence of elliptically polarized states with two
points in the core of the wall at which the state becomes linearly
polarized ($s_3=0$). The change in polarization state still occurs
for $s_1 \simeq 0$ because being $c_x^i=0$, still $\Gamma=1$. The
representative point moves now in the Poincare sphere from a point
close to the north pole to the vicinity of the south pole crossing
the equator, and back to the original point along the other side
of the sphere crossing again the equator. An opposite sense of
motion along the sphere would correspond to an opposite chirality
of the wall.

We note that a quantitative precise description of the variation
of the Stokes parameters can be generally given by invoking the
relation, that follows from symmetry considerations, $A_x \simeq i
A_y^*$ point wise along the wall. The Stokes parameters as
functions of the spatial variation of the phase $\psi_x(x)$, are
then given by (substitute $A_x=i A_y^*$ into eqs. (\ref{stokes}):
\begin{eqnarray}
s_1(x) &\simeq& 0 \nonumber \\
s_2(x) &\simeq& sin(2 \psi_x(x)) \nonumber \\
s_3(x) &\simeq& -cos(2 \psi_x(x))
\label{pol-wall}
\end{eqnarray}

We next turn to consider the dynamics of the domain
walls, which is also useful to determine the transition between Ising
and Bloch walls. Isolated Ising walls in 1D are stable and remain
stationary. The dynamics of 1D BWs depends critically on the
values of the cavity decay rates and the detuning. For $\gamma_x
\Delta_x=\gamma_y \Delta_y$, 1D BWs do not move: they are stable
stationary interfaces between equivalent uniform domains. A
similar situation takes place in the potential limit of the 1D
parametrically forced Complex Ginzburg Landau Equation (PCGLE)
where stationary BWs have been found analytically
\cite{Coullet90,Tutu97}. On the contrary, for $\gamma_x \Delta_x
\ne \gamma_y \Delta_y$ walls of different chirality move in
opposite directions, as it also happens outside the potential
limit of the (PCGLE). The velocity of the resulting BW depends on
the value of the parameters, in particular it depends strongly on
$c_{x,y}$. In figure \ref{figspeed} the velocity of BWs as a
function of $c_x$ (real), for selected values of the other
parameters, is shown as it results from  numerical solutions. For
small $c_x$ BW are not stable because we are outside the locking
regime ( eq. (\ref{condition}) is not satisfied); for larger
values the velocity decreases by increasing $c_x$ and finally it
vanishes. The vanishing of the velocity identifies the transition
point in which a BW decays into an IW. This transition is
continuous, i.e. the amplitude of the wall solutions, that is
almost constant for BWs for small $c_x$, shows larger and larger
variations as the transition is approached. The amplitude of the
signal field (and also the idler) become small at the core of the
wall and it eventually reaches a zero. At this point BWs and IWs
exchange their stability and only standing IWs are found beyond
that critical value. The conclusion is that in the regime of the
parameter space for which BWs are stable, IWs are unstable and
vice versa.

In addition to the transition from Bloch to Ising type, the
strength of the direct polarization coupling coefficients
$c_{x,y}$ also controls the wall width. Moving well within the
locking regime the width becomes small, while it diverges  as $c_{x,y}
\rightarrow 0$.
\newline
We finally mention that other forms of stable localized structures
or domain walls are sustained in this system, although they do not
appear spontaneously when starting from a weakly perturbed trivial
steady-state (\ref{trivial}) and a pump value above threshold.
They can be formed far above threshold mediated by the unstable
phase locked solutions of higher threshold discussed in Sect. III.
A first example, shown in Fig. \ref{fighomoc}a is a stable 1D
localized structure in the background of a + in-phase solution.
This is structure is generated from a step-like initial condition,
shown in Fig. \ref{fighomoc}b, which connects the stable + in-phase
solution with the unstable + out-of-phase solution. The dynamical
evolution of this unstable configuration leads to the localized
structure. A second example shown in Fig. \ref{fighomoc}c is a new
kind of domain wall which connects, as in the Ising and Bloch
walls discussed above, the equivalent + and - solution of the
phase locked solutions of lowest threshold. The difference with
our previous Ising walls consists in a more complicated structure
of the core of the wall. In this case the field $A_x$ vanishes at
three points in the core of the wall. This new type of domain wall
emerges dynamically from a similar initial condition that the one
shown in Fig. \ref{fighomoc}b, but changing the sign of the
unstable solution of the step: The initial step like condition
connects the stable + in-phase solution with the unstable -
out-of-phase solution.

\section{CHARACTERIZATION OF 2D BLOCH WALLS}

In our 2D numerical solutions \cite{numerical} with random initial
conditions around the trivial unstable solution we also observe
the spontaneous formation of BWs which are now lines in the
transverse plane. These walls evolve dynamically as described
below. A main new feature of BWs in 2D is that the domain walls
can emerge with an opposite chirality in different spatial regions
along the wall. The points on the wall where the chirality changes
sign are singular points: At these points the phases of the signal
and idler field are not defined and the amplitudes go to zero,
i.e. they can be classified as topological defects. The BW, in
that particular point, actually degenerates into an IW. A snapshot
of typical transient transverse patterns generated by the Bloch
walls is shown in figure \ref{fig2d-bw} for the component $A_x$.
The two equivalent + and - phase locked homogeneous solutions are
represented in Fig. \ref{fig2d-bw} a) by regions of different
intensity on a gray scale ($A_x^{\pm}$). Likewise, the segments 
along the walls
of different chirality are represented respectively by black or
white segments ($B_{\pm}$) in figure \ref{fig2d-bw} a). The
defects, where the changes of chirality take place can be observed
as black dots in the intensity field (figure \ref{fig2d-bw} b)).
Note that, except for the defects, the field intensity is almost
constant in the domains and only slightly modulated close to the
core of the wall. The phase field, shown in figure \ref{fig2d-bw}
c), demonstrates that phase defects with topological charge $\pm
1$ occur at the points of change of chirality on the wall. An
expanded vision of the amplitude of the field in the vicinity of
one of these defects is shown in Fig. \ref{fig3d}. The corresponding
snapshot for the field $A_y$ shows the same structure with walls
and defects at the same positions as for $A_x$.

There are different effects determining the dynamics of BWs in 2D.
A first mechanism is the one found in 1D, related with detuning
and damping coefficients. However, in 2D the dynamics is also influenced
by the curvature of the walls and the presence of field defects.
In what follows two main regimes are identified. In a first
regime, called of domain growth, curvature effects dominate. This
leads to the complete disappearance of all fronts, all defects and
all domains except one, the final state being an homogeneous
solution. The second is characterized by a persistent dynamics in
which defects are deeply stable objects, while domains and walls
are continuously generated and annihilated.

\subsection{Regime of domain growth}
For $\gamma_x \Delta_x=\gamma_y \Delta_y$, flat 2D BWs are stable
structures. This corresponds to the fact that 1D walls, for the
same values of the parameters, do not move. The transient dynamics
is then ordering process mainly controlled by the curvature
effects in which the walls evolve by reducing curvature. This
leads to the growth of one of the two equivalent solutions at the
expenses of the other and the annihilation of all the defects.
This process is shown in figure \ref{figgrowth}, where snapshots
at different times are shown. The images of the upper row show the
intensity of polarization component $A_x$, while in the pictures of
the lower row, the real part of the same field is presented.
Slowly all the defects and walls disappear and the final state is
a homogeneous phase-locked solution.
\newline
In this regime the normal velocity of the fronts is determined by
the local curvature of the wall. This is demonstrated by the
result of figure \ref{figvel} where the evolution of the square
radius of a domain surrounded by a circular BW has been
determined. The result is a growth law $R(t) \simeq t^{1/2}$
characteristic of curvature driven domain growth \cite{rafa}.
Similar structures, annihilation dynamics and dynamical exponents
for the growth law of BWs domains, have been reported in the
description of the ordering process of a non conserved anisotropic
XY-spin system in 2D \cite{Tutu97}.

\subsection{Regime of persistent dynamics}

More peculiar of our system is the regime found for $\gamma_x
\Delta_x \ne \gamma_y \Delta_y$. Let's recall that in this regime
1D optical BWs of different chirality move in opposite directions
while IWs (characterized by a point of zero amplitude) do not
move. The dynamics in 2D is reminiscent of this behavior; in fact
the defects are notably stable as the corresponding 1D IWs, while
BWs of different chirality move in opposite directions. The
combination of these two effects is that BWs spiral around the
defects; such phenomenon was also observed for BWs in other
physical systems \cite{frisc94}. The spiraling dynamics of an
isolated defect is shown in Fig. \ref{figspiral} obtained with
a flat-top profile for the pump beam. The stability of the defects,
with these physical boundary conditions is remarkable. This is an
important result because stabilization of optical vortices has
been always critical in lasers and nonlinear optical systems \cite{stal99}.
When many of these defects arise spontaneously from random initial
conditions around the trivial unstable solution, the system
becomes trapped in a complicated state of persistent dynamics in
which an homogeneous state is never reached. An understanding of
such persistent dynamics is easier for particular initial
conditions. In the example shown in Fig. (\ref{figcoll}) the initial
condition is a flat BW with equally spaced defects along the wall.
This gives rise to segments of different chirality along the wall
that start moving in opposite directions. The net result, after
collision of spiraling BWs of the same chirality is the periodic
emission, in each direction, of BWs of alternating chirality. The
defects remain stable and walls are re-generated by the spiraling
process. The dynamical process that we have just described becomes
fuzzy close to the Ising-Bloch transition. In such regime of
parameters the amplitude of the field becomes very small at the
core of the wall and the clear distinction between point defects
and wall segments of different chirality is lost.

\section{2D ISING WALLS}

Beyond the transition from BWs to IWs the latter appear
spontaneously separating spatial domains of the + and - solutions,
but their dynamics and the asymptotic state of the system still
depends strongly on the strength of the direct polarization
coupling. As in the case of BWs, we find for IWs two main regimes:
a regime of domain growth and one of labyrinthine pattern
formation.

For coupling values close to the Bloch-Ising transition, a flat
Ising wall is stable. The transient dynamics is then controlled by
the curvature of the walls. Curvature reduction leads to domain
growth much in the same way hat we already described for BWs in
the corresponding regime. In figure \ref{figcoars} a typical time
series of snapshots of the transverse $A_x$ field is shown. The
final asymptotic state is homogeneous. Similar dynamics has
been already reported in type-I DOPO above threshold
\cite{oppo01,tril97,oppo98}, the only difference being the
vectorial character of the field in this case.

Far from the Bloch-Ising transition and moving deeper into the
locking regime transverse labyrinthine patterns are spontaneously
formed in the system starting from a random perturbation of the trivial
unstable solution \cite{rafa,tara98}. Snapshots of the evolution
of a pattern of this type is shown in intensity and real part in
figure \ref{figlabir}. Note that the time evolution of
labyrinthine patterns is very slow.
The creation of a labyrinthine pattern stems from the
fact that, in this regime of parameters, flat Ising walls (i.e.
with no curvature) are modulationally unstable. Roughly speaking
the finger growth is associated with a band of modulational
frequencies of the front curvature that tend to increase their
curvature. This is reminiscent of what has been reported for intra
cavity second harmonic generation \cite{lede98} and vectorial Kerr
resonators \cite{rafa} . In order to illustrate the modulational
instability, the evolution of an initially perturbed Ising, flat
wall is shown in figure \ref{figmi}.

\section{CONCLUSION}

In conclusion we have demonstrated that Bloch walls can be found in
nonlinear optical systems, in particular in type-II optical parametric
oscillators. They appear when there exists a small birefringence and/or
dichroism of the cavity along axes that do not coincide with those of the
nonlinear crystal. These effects introduces a linear coupling between the
signal and the idler which causes self phase-locking of the two fields.
There exist two possible steady-state solutions, characterized by a phase
shift of $\pi$ radiants of both polarization components and thus different
domains spontaneously form in which one or the other solution is selected.
Separating walls can be either of the Bloch or the Ising type depending on
the strength of the coupling coefficient. For small values Bloch walls are
stable and appear spontaneously above a predicted threshold out of a random
perturbation of the trivial steady-state. 
Bloch walls have been characterized both in one and two dimensions.
In one dimension a physical interpretation of Bloch walls is given in terms
of polarization variations that connect two homogeneous states that
represent the same state of polarization. The chirality is instead related
to the ellipticity variations. Multiple hump Ising walls have also been
found starting from particular initial condition.
In two dimension, Bloch wall peculiarity is the possibility
of having wall sections of different chirality, i.e. where the phase
rotates in two possible ways, clockwise or counterclockwise in the complex plane. Where
chirality changes sign the phase has defects, where the field amplitude is zero, and
the wall degenerate into an Ising one. Two dynamical regimes, that depend on the
decay rates and the detunings, are found: in the first one, the wall
dynamics is dominated by the curvature and a final homogeneous state is
reached; in a second regime, the walls spiral around stable defects and a
persistent creation and annihilation of fronts is observed. The transition
from Bloch to Ising walls has been also observed when the linear coupling
strength is increased; transition is characterized by larger and larger
variations of the amplitude of the field close to the core of the wall
and the fact that walls stop moving. Ising wall dynamics has been also considered,
in particular their curvature modulation instability, that
leads to the creation of a labyrinthine pattern.

This work has been partially supported by the Spanish MCyT project
BFM2000-1108 and by the European Commission project QSTRUCT
(FMRX-CT96-0077). The authors acknowledge clarifying discussions
with G-L Oppo.

\section{APPENDIX}

In this appendix the derivation of eqs. (\ref{master})
is presented. The main difference with respect to previous
derivations of mean filed equations for OPOs is the inclusion of
the effects of dichroic and birefringent mirrors. In order to
simplify the model, let us assume that only one, out of the four
mirrors of the ring cavity, is birefringent and dichroic. This
means that, in a proper system of orthogonal axes, the matrix $M$,
that represents the transformation of two orthogonally polarized
component of a beam, in the Jones formalism \cite{phot}, is:
\begin{equation}
M=
\left(
\begin{array}{cc}
   r_1 e^{-i\psi_1} & 0 \\ 0 & r_2 e^{-i\psi_2}
\end{array}
\right)
= r e^{-i\sigma}
\left(
\begin{array}{cc}
   (1+p) e^{i\delta} & 0 \\ 0 & (1-p) e^{-i\delta}
\end{array}
\right)
\end{equation}
where $r=(r_1+r_2)/2, p=(r_1-r_2)/(2r), \sigma=-(\psi_1+\psi_2)/2$ and
$\delta=(\psi_2-\psi_1)/2$.
The dichroism, i.e. the different reflectivity of the mirror for different
polarizations ($r_1 \neq r_2$), implies that $p \neq 0$, while the
birefringence is represented by the fact that $\psi_1 \neq \psi_2$, i.e. the
phases of the reflected components undergo a different change
($\delta \neq 0$).
\newline
A first important remark to make is that the mirror anisotropy
axes might not coincide with the axes of the nonlinear crystal,
that is also birefringent in order to realize the phase matching between
the
pump and the generated fields. In other words the anisotropic
crystal has its own preferred polarization axes that can be
rotated by an angle $\phi$ with respect to the mirror principal
axes (those for which the mirror matrix is diagonal). Therefore,
when passing from the propagation in the cavity axes reference
frame to that in the crystal a rotation, represented by the matrix
\begin{equation}
R=
\left(
\begin{array}{cc}
   cos \phi & -sin \phi \\ sin \phi & cos \phi
\end{array}
\right),
\end{equation}
has to be applied. At the end of the propagation in the nonlinear
medium an inverse transformation ($R^{-1}$) is needed to restore
reference frame of the cavity axes.
\newline
The generic $n+1$ round trip in the cavity for the signal and
idler vector field of the previous step n ($\vec{E}_n$) is then
represented by the following transformation (all mirrors except
the last one are perfectly reflective, i.e. their matrices are all
equal to the identity matrix):
\begin{equation}
\vec{E}_{n+1}=M \; R^{-1} \; f(R \vec{E}_n)
\label{round}
\end{equation}
where $f(\cdot)$ is the result of the propagation inside the nonlinear
medium. A similar
 formula can be written also for the pump vector field
$\vec{F}_n$
that, in principle, has both polarization components. Usually
one
component does not participate in the nonlinear dynamics (it is
not phase-matched) and it is neglected a priori. In this case, due
to polarization coupling, it is included in the model although the
final result is that, under not very restrictive hypotheses, its
effects can be neglected.
\newline
The output of the function $f(\cdot)$ is the vector field as it results after
the integration of a set of nonlinearly coupled propagation equations, i.e.
it involves products of $\vec{E_n}$ and $\vec{F}_n$. Hereafter only the signal
and idler vector field $\vec{E_n}$ equations will be considered; similar
calculations can be repeated for the pump components.
\newline
Let's remark that the optical carrier frequency of each component
has been already removed (i.e. envelope equations \cite{opos} are
searched for) and that the carrier frequencies of all waves are
determined by three conditions: phase-matching and energy
conservation of the nonlinear interaction and the condition of
resonance due to the cavity. It has been demonstrated
\cite{fabr93} that, for a type II OPO, there are several
signal-idler pairs of oscillation frequencies that can satisfy
these conditions. Among these solutions there is also the case of
quasi, or totally, frequency degenerate signal and idler. This is
also verified experimentally by the fact that type-II OPO's,
differently from type-I, can be smoothly tuned through frequency
degeneracy \cite{Mason98,opos}. Let us remark, finally that for
type II OPOs signal and idler are always polarization
non-degenerate.
\newline
The boundary condition to impose into eq. (\ref{round}) for
steady-state operation of the OPO is that the round trip
transformation coincides with the identity, i.e.
$\vec{E}_{n+1}=\vec{E}_n$. Let us define
$\vec{A}(L)=f(R\vec{E}_n)$ the vector field at the output of the
crystal of length L (that is also the cavity length for the
completely filled cavity) and
$\vec{A}(0)=R\vec{E}_n=R\vec{E}_{n+1}$ the field at the crystal
input; by multiplying the left and right hand side of eq.
(\ref{round}) by $R$ (on the left) and substituting previous
definitions the following equation is found:
\begin{equation}
\vec{A}(0)=R \; M \; R^{-1} \; \vec{A}(L)
\end{equation}
Let us now set $\vec{A}'(z)=H(z)\vec{A}(z)$ such that
\begin{equation}
\vec{A}'(0)=\vec{A}(0)\; , \; \vec{A}'(L)=\vec{A}'(0)
\label{boundary}
\end{equation}
the second condition imposing the periodicity after a round trip.
The general form of the matrix $H(Z)$ satisfying (\ref{boundary}) is:
\begin{equation}
H(z)=
\left(
\begin{array}{cc}
   e^{h_{xx}z} & e^{h_{xy}z}-e^{-h_{xy}z} \\  e^{h_{yx}z}-e^{-h_{yx}z} & e^{h_{yy}z}
\end{array}
\right)
\end{equation}
This matrix extends the scalar transformation used by Lugiato and Oldano in
their original paper \cite{lugi88}, devoted to the study of stationary spatial
patterns in optical systems with two-level atoms, to a vectorial case. It is easy
to verify that $H(0)$ is the identity matrix while $H(L)=RMR^{-1}$ and thus all
the elements of the matrix $h_{ij}$ can be explicitly calculated:
\begin{eqnarray}
h_{xx} &=& \frac{1}{L} \{ ln(r)-i\sigma+ln[(1+p \, cos 2 \phi) cos \delta+
i (p+cos 2 \phi) sin \delta] \} \nonumber \\
h_{yy} &=&\frac{1}{L} \{ ln(r)-i\sigma+ln[(1-p \, cos 2 \phi) cos \delta+
i (p-cos 2 \phi) sin \delta] \} \nonumber \\
h_{xy} &=& h_{yx}=\frac{1}{L} \{ ln[ (p \, cos \delta + i \, sin \delta) sin 2\phi +
\sqrt{(p \, cos \delta + i \, sin \delta)^2 sin^2 2 \phi + 4}] -ln(2) \}
\label{haches}
\end{eqnarray}
The evolution of the field in the crystal is governed by equations of the
type:
\begin{equation}
\partial_z \vec{A} = {\cal L}(\vec{A}) + {\cal N}(\vec{A},\vec{B})
\end{equation}
where $\vec{B}=[B_x, B_y]=R \vec{F}_n$ represents the pump vector
field. The linear term is
\begin{equation}
{\cal L}(\vec{A})=
\left(
\begin{array}{cc}
   \frac{i}{2k_x} \nabla^2 - \frac{1}{v_x} \partial_t & 0 \\
0 & \frac{i}{2k_y} \nabla^2 - \frac{1}{v_y} \partial_t
\end{array}
\right) \vec{A}
\end{equation}
and includes the diffraction ($k_{x,y}$ are the longitudinal
wave vectors of signal and idler, $\nabla^2$ is the spatial transverse
Laplacian operator) and the phase velocity mismatch ($v_{x,y}$ are the phase
velocity respectively of signal and idler, $\partial_t$ is the differential
operator with respect to time). The nonlinear operator is
\begin{equation}
{\cal N}(\vec{A},\vec{B})= i K  B_x
\left(
\begin{array}{cc}
   0 & 1 \\
1 & 0
\end{array}
\right) \vec{A}^*
\end{equation}
where $K$ is the nonlinear coefficient.
Since $\vec{A}=H^{-1} \vec{A}'$ the evolution of the field $\vec{A}'$ in
the crystal can be determined from:
\begin{equation}
\partial_z \vec{A} =  (\partial_z H^{-1}) \vec{A}' + H^{-1} \partial_z \vec{A}'
\end{equation}
that finally yields:
\begin{equation}
\partial_z \vec{A}' + H (\partial_z H^{-1}) \vec{A}' =  H {\cal L} H^{-1}
\vec{A}' + H {\cal N} (H^*)^{-1} \vec{A}'
\end{equation}
The matrix $H$ is known and so all the terms of the last equation
can be explicitly calculated; in particular, by following the
guidelines of ref. \cite{lugi88}, if $p$ and $\delta$ are small
parameters (of the order of the transmittivity T=1-r of the
mirror) one can calculate a set of first order (in T) equations by
expanding all coefficients up to this order. After long but
straightforward calculations, the final result is a set of coupled
equations for the vector $\vec{A}'=[A_x, A_y]^T$, that still
contains both $\partial_z$ and $\partial_t$ operators, but
includes the boundary conditions. It is worth to write these
equation because the coefficient of each term appears explicitly
written in terms of physical parameters:
\begin{eqnarray}
\partial_t A_x + v_x \partial_z A_x = h_{xx} v_x A_x + \frac{v_x}{L} (p+i \delta)
sin 2 \phi A_y + i \frac{v_x}{2 k_x} \nabla^2 A_x + i v_x K A_y^* B_x  \nonumber \\
\partial_t A_y + v_y \partial_z A_y = h_{yy} v_y A_y + \frac{v_y}{L} (p+i \delta)
sin 2 \phi A_x + i \frac{v_y}{2 k_y} \nabla^2 A_y + i v_y K A_x^* B_x
\label{prop}
\end{eqnarray}
By exploiting the single longitudinal mode approximation, which is quite a
good one for continuous wave OPO's, the longitudinal spatial dependence can
be finally removed and the equations describing the time evolution are
exactly eqs. (\ref{master}).
The coefficients of eqs. (\ref{master}) can be found from eqs. (\ref{prop})
once the expressions (\ref{haches}) are substituted. They read:
\begin{eqnarray}
\gamma_{x,y}=\frac{v_{x,y} (T \pm p \, cos 2 \phi)}{L} \nonumber \\
\Delta_{x,y}=\frac{v_{x,y} (\sigma \pm \delta \, cos 2 \phi)}{T \pm p \, cos 2 \phi}
\nonumber \\
\alpha_{x,y}=\frac{L}{2 k_{x,y} (T \pm p \, cos 2 \phi)} \nonumber \\
c_{x,y}=\frac{p + i \delta}{T \pm p \, cos 2 \phi} sin 2 \phi
\label{coeffs}
\end{eqnarray}
where the plus (minus) sign applies for the x-polarized (y-polarized)
component.
Note that all coefficients can be different because of the crystal
birefringence ($v_x \neq v_y$), the cavity birefringence and/or dichroism
($p \neq 0, \delta \neq 0$). Finally the nonlinear coefficients is defined
as
\begin{equation}
K_0 = \frac{K L}{T}
\label{nlcoeff}
\end{equation}
Actually it is also slightly different for the two polarization, but this
difference has been neglected because it is only due to the mirror
dichroism; all the other coefficients have larger differences because two
effects (crystal and cavity birefringence) contribute.
\newline
Regarding the coupling coefficients some special cases are worth
to be remarked in relation to what is discussed in other sections
of this paper. For example when $cos 2 \phi=0$, the special
condition $c_x=c_y$ (complex) is obtained; in this case $\phi =
\pm \pi/4, \pm 3 \pi/4$, $sin 2 \phi=\pm 1$ and therefore the
linear coupling among polarization is maximized in modulus. Other
two special cases are the purely birefringent mirror $p=0$, for
which $c_x=c_y$ are purely imaginary, and the purely dichroic
mirror that yields $\delta=0$ and thus purely real $c_{x,y}$.

\newpage

\centerline{\bf FIGURES}

\begin{figure}
\caption{Bifurcation diagram of the homogeneous solutions 
(solid curves stable, dashed curves unstable). Parameters are $K_0=1$, 
$c_y=c_x=0.05 \, (1+i)$, $c_x'=c_y'=0.1$, $\Delta_x'=\Delta_y'=0$, $\Delta_x=0.01$, $\Delta_y=0.03$.}
\label{figbifur}
\end{figure}


\begin{figure}
\caption{Threshold of instability $E_c$ 
for the trivial stationary solution (eq. (\ref{trivial})): solid 
(dashed) curve is the threshold of the in- (out-of-) phase solution. 
Here $c_y=c_x$ and the others parameters are the same of figure 
\ref{figbifur}.}
\label{figthresh}
\end{figure}


\begin{figure}
\caption{Numerical solution of eqs. (\ref{master}) in one spatial 
dimension showing and example of an Ising wall: a) Solid (dotted) curve 
is the real (imaginary) part of the field $A_x$ as a function of the
transverse coordinate; b) the same wall is represented in the complex 
plane ($Re(A_x),Im(A_x)$) in solid line. Dotted curve is the 
Ising wall associated to the field $A_y$. The parameters are $\gamma_y=\gamma_y'=1$, $\gamma_x=\gamma_x'=1.001$,
$\Delta_x'=\Delta_y'=0$, $\Delta_x=0.01$, $\Delta_y=0.03$, $\alpha_x'=\alpha_y'=0.125$, $\alpha_x=\alpha_y=0.25$, $K_0=1$, $E_0=1.25$, $c_x'=c_y'=0.01(1+i)$ and $c_x=c_y=0.082$.}
\label{figiw-1d}
\end{figure}


\begin{figure}
\caption{Numerical solution of eqs.(\ref{master}) in one spatial 
dimension, showing a Bloch wall. a) Solid (dotted)  curve is the 
real (imaginary) part of the field $A_x$ as a function of the transverse coordinate; b) the same wall is represented in the complex plane ($Re(A_x),Im(A_x)$) in solid line. Dotted curve is the Bloch wall associated 
to the field $A_y$. Parameters are the same of figure \ref{figiw-1d}, except $c_x=c_y=0.01$. }
\label{figbw-1d}
\end{figure}


\begin{figure}
\caption{Variation of the polarization along a Bloch wall, represented
by means of the Stokes parameters (\ref{stokes}) as they result 
from a numerical solution: $s_1(x)$ (solid curve), $s_2(x)$ (dashed), 
$s_3(x)$ (dotted). a) The Bloch wall separates linearly polarized states.
Parameters are the same of figure \ref{figiw-1d}, except 
$\gamma_x=\gamma_x'=1$, $\Delta_x=\Delta_y=0.03$, $c_x'=c_y'=0.01$, $c_x=c_y=0.02$; b) The same representation for the Bloch wall of Fig. \ref{figbw-1d} which separates elliptically polarized states.}
\label{figstokes}
\end{figure}


\begin{figure}
\caption{Velocity of a BW as a function of 
$c_x^r$ ($c_x^i=0$) as given by numerical simulations of eqs. (\ref{master}) 
in the regime $\gamma_x \Delta_x \neq \gamma_y \Delta_y$. The parameters 
are the same of figure \ref{figiw-1d}, except the values of $c_y^r=c_x^r$. 
For the last three points in the figure (corresponding to Ising walls) the velocity is zero. The dotted line indicates the onset of the locking regime.}
\label{figspeed}
\end{figure}


\begin{figure}
\caption{Examples of multiple humps stable solutions: in a) the field comes 
back to the initial state;  b) initial condition used in (a). Here $Re(A_x)=Im(A_x)$ for $x<0$; c)  IW with two crossings of the zero
amplitude point. Parameters are $\alpha_x = 0.25375$, $\alpha_y = 0.24625$
$\alpha_x'=\alpha_y'=0.125$, $\Delta_x'=\Delta_y'=0$, $\Delta_x=\Delta_y=0.02$,
$\gamma_x'=\gamma_y'=1$, $\gamma_x=0.985$, $\gamma_y=1.015$, $c_x'=c_y'=0.01$,
$c_x=c_y=0.2+i \, 0.02$.}
\label{fighomoc}
\end{figure}


\begin{figure}
\caption{A snapshot at time t=1600 of the field $A_x(x,y,t)$, spontaneously generated from random initial conditions close 
to the trivial steady-state. Figure a), b) and c) show respectively the real part, the intensity and the phase. Black and white segments in the walls ($B_{\pm}$) in a) correspond to opposite sense of rotation of the phase (chirality). The black points in b) are the defects, where signal and idler amplitude vanishes. The parameters are the same of figure \ref{fighomoc}, 
except $\Delta_x=0.01$, $\Delta_y=0.03$,  $c_x'=c_y'=0.025 (1-i/2)$ and $c_x=c_y=0.02(1+i)$.}
\label{fig2d-bw}
\end{figure}


\begin{figure}
\caption{Amplitude of the $A_x$ field as a function of the transverse coordinates $(x,y)$ in the vicinity of a defect. The parameters are 
the same of figure \ref{figiw-1d}, except $\gamma_x=\gamma_x'=1$, $c_x=0.02(1+i/2)$, $c_y=0.02$.}
\label{fig3d}
\end{figure}


\begin{figure}
\caption{The domain growth regime ($\gamma_x \Delta_x= \gamma_y \Delta_y$)
is presented by means of snapshots at different times: a)$t=200$; b) $t=600$; 
c) $t=1000$; d) $t=2000$. The upper row shows the evolution of the intensity 
of $A_x(x,y,t)$ while in the lower row the real part of the same field is 
shown. The initial condition is random and the parameters are the same of 
figure \ref{figiw-1d}, except $\Delta_y=0.01002$, $\gamma_1=1.002$, $c_x=c_y=c'_x=c'_y=0.02 (1+i)$. }
\label{figgrowth}
\end{figure}


\begin{figure}
\caption{The asterisks shows the time evolution of the square 
radius $R^2$ of a circular BW domain, in the growth domain regime, as they result form the numerical solution. The solid line is a linear interpolation 
of numerical data. Parameters are the same of figure \ref{figgrowth}, except
$c'_x=c'_y=0.01 (1+i)$ and $c_x=c_y=0.02$. }
\label{figvel}
\end{figure}


\begin{figure}
\caption{Time evolution of one spiral of BWs around a defect.
In the left column the amplitude of $A_x(x,y,t)$ is represented while 
in the right column the real part of the same field is shown. a) $t=560$; b) $t=1120$. The parameters are the same of figure \ref{fig3d}.}
\label{figspiral}
\end{figure}


\begin{figure}
\caption{Time evolution of a BWs, spiraling around an array of defects,
imposed as an initial condition: a) t=0; b) t=1000; c) t=1150; d) t=1550, 
in the left column the amplitude of $A_x(x,y,t)$ is represented while 
in the right column the real part of the same field is shown. The parameters 
are the same of figure \ref{fighomoc} except $\Delta_x=0.01$, $\Delta_y=0.03$, $c_x=c_y=0.025$ and $c_x'=c_y'=0.01$.}
\label{figcoll}
\end{figure}


\begin{figure}
\caption{Snapshot at different times, observed in the 
coarsening regime: a) $t=300$; b) $t=900$; c) $t=1900$; d) $t=3000$.
The upper row presents the intensity pattern and 
the lower row $Re(A_x)$. The initial condition is random and the values 
of the parameters are the same of figure \ref{figiw-1d}, except:
$c_x=c_y = 0.09$.}
\label{figcoars}
\end{figure}


\begin{figure}
\caption{Formation of a labyrinthine pattern in the regime for which IWs are stable but their curvature is modulationally unstable: the upper row shows the intensity pattern and the lower row $Re(A_x)$ for a) $t=75$; b) $t=1500$. The initial condition is random and the values of the parameters are the same of figure \ref{figcoars}, except: $\Delta_y=0.01001$, $c_x=c_y=c'_x=c'_y= 0.3 \, i$. }
\label{figlabir}
\end{figure}


\begin{figure}
\caption{Modulational instability of an initially flat IW: upper row shows $|A_x(x,y,t)|^2$ and the lower row the real part of the same field.
In this regime 1D Ising front are stable. The values of the parameters are
 the same of figure \ref{figlabir}.}
\label{figmi}
\end{figure}

\newpage

\centerline{\psfig{figure=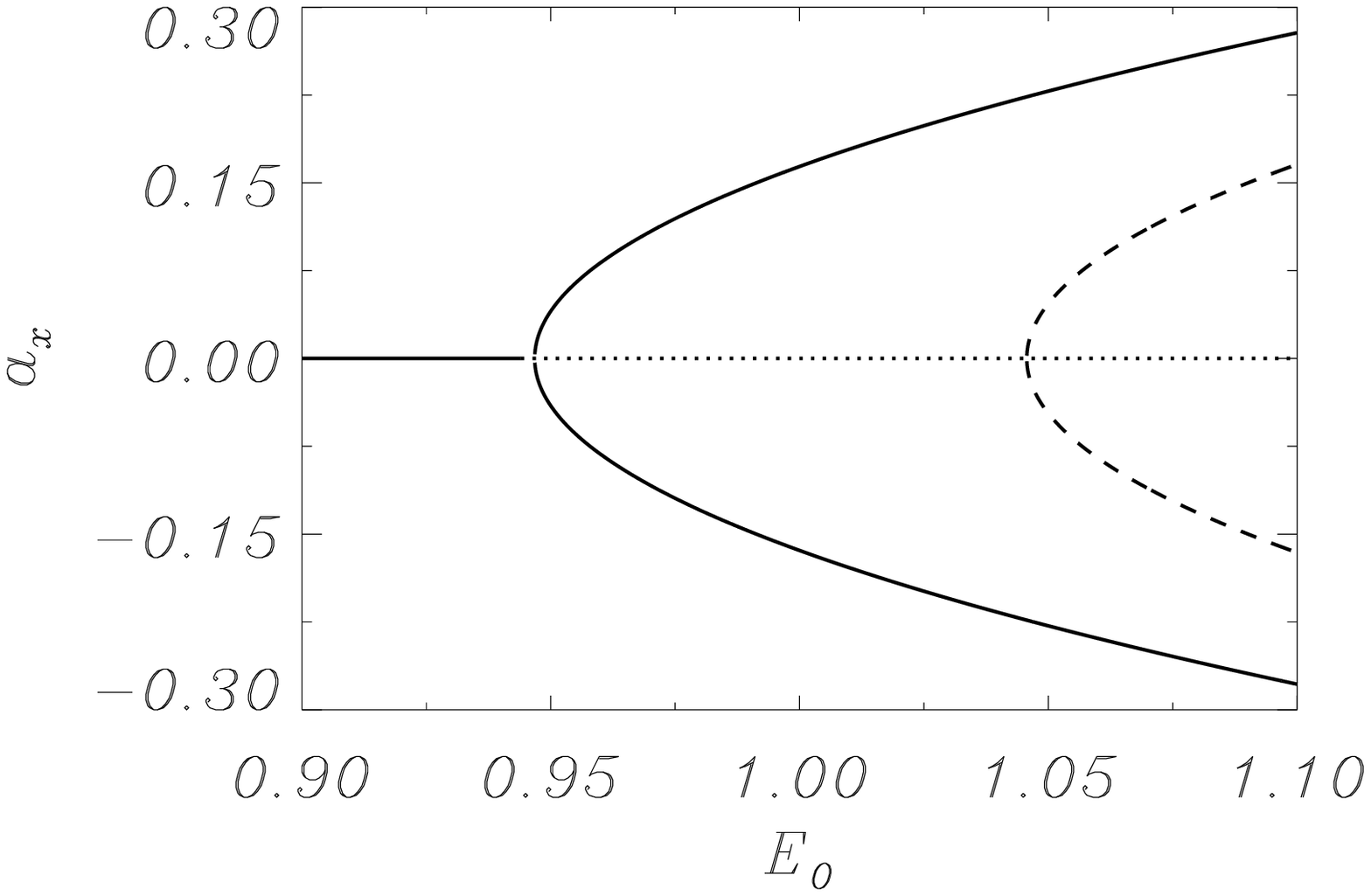,width=5in}}
\centerline{Figure 1}
\centerline{\psfig{figure=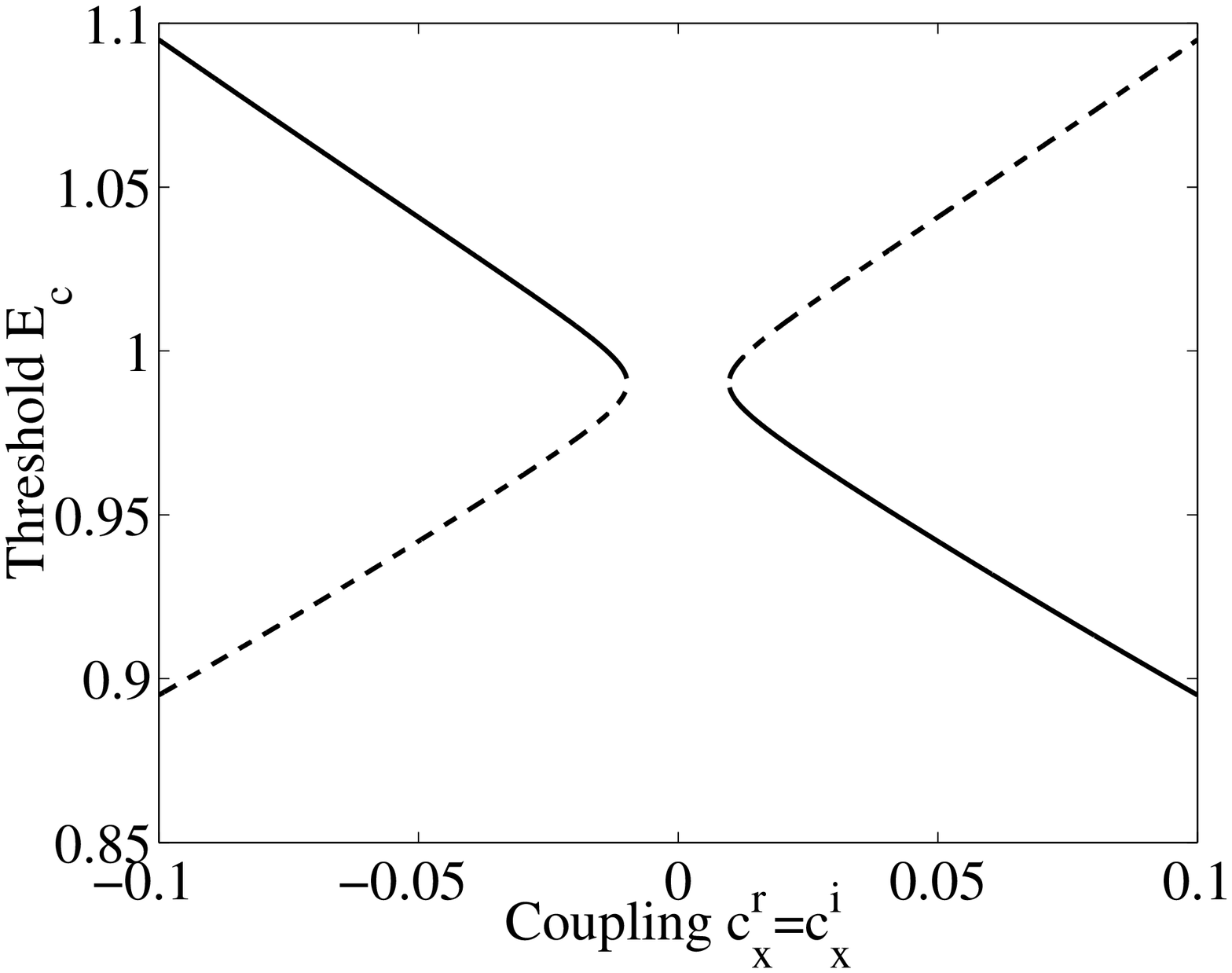,width=5in}}
\centerline{Figure 2}

\newpage

\centerline{\psfig{figure=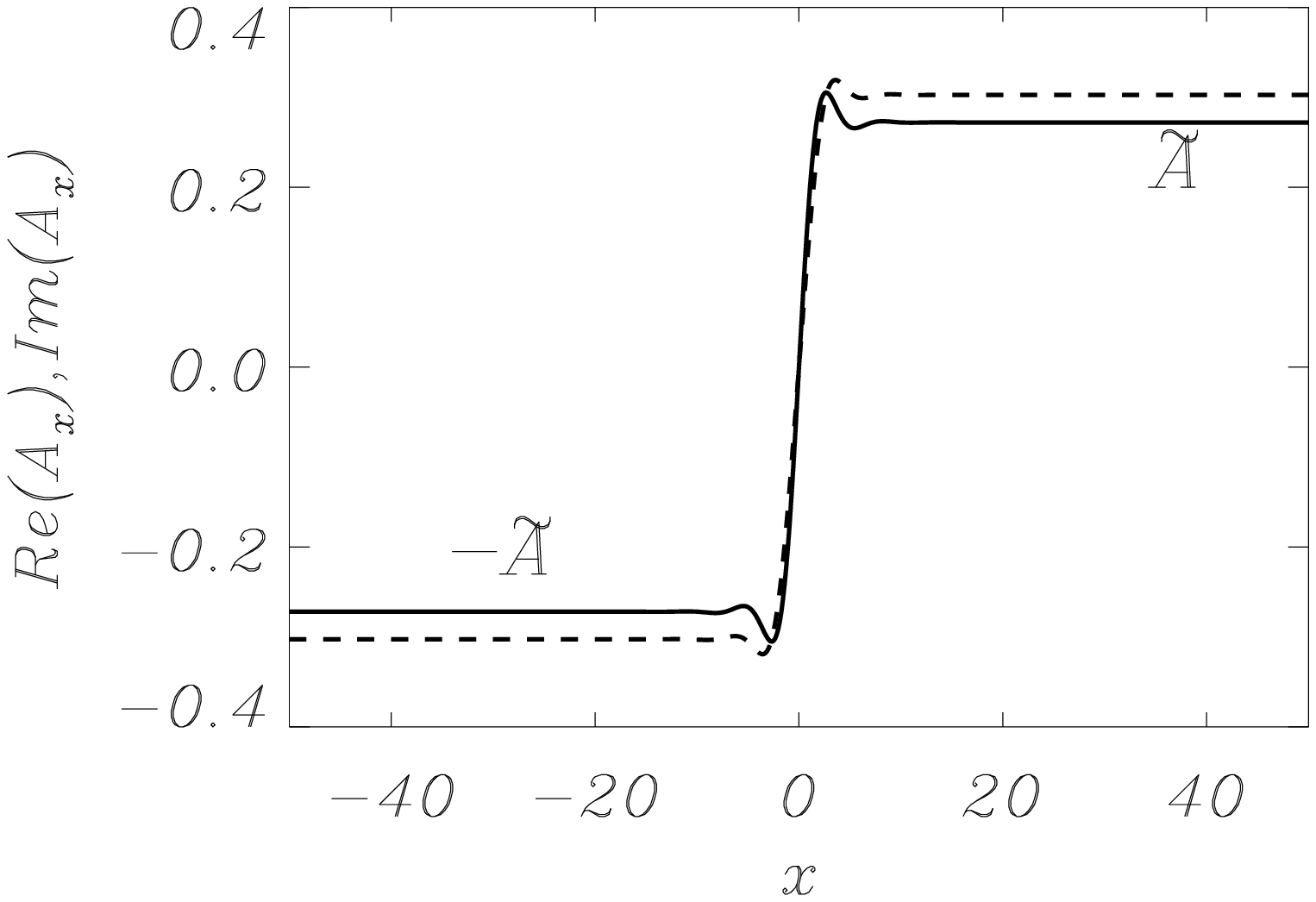,width=5in}}
\centerline{Figure 3a}
\centerline{\psfig{figure=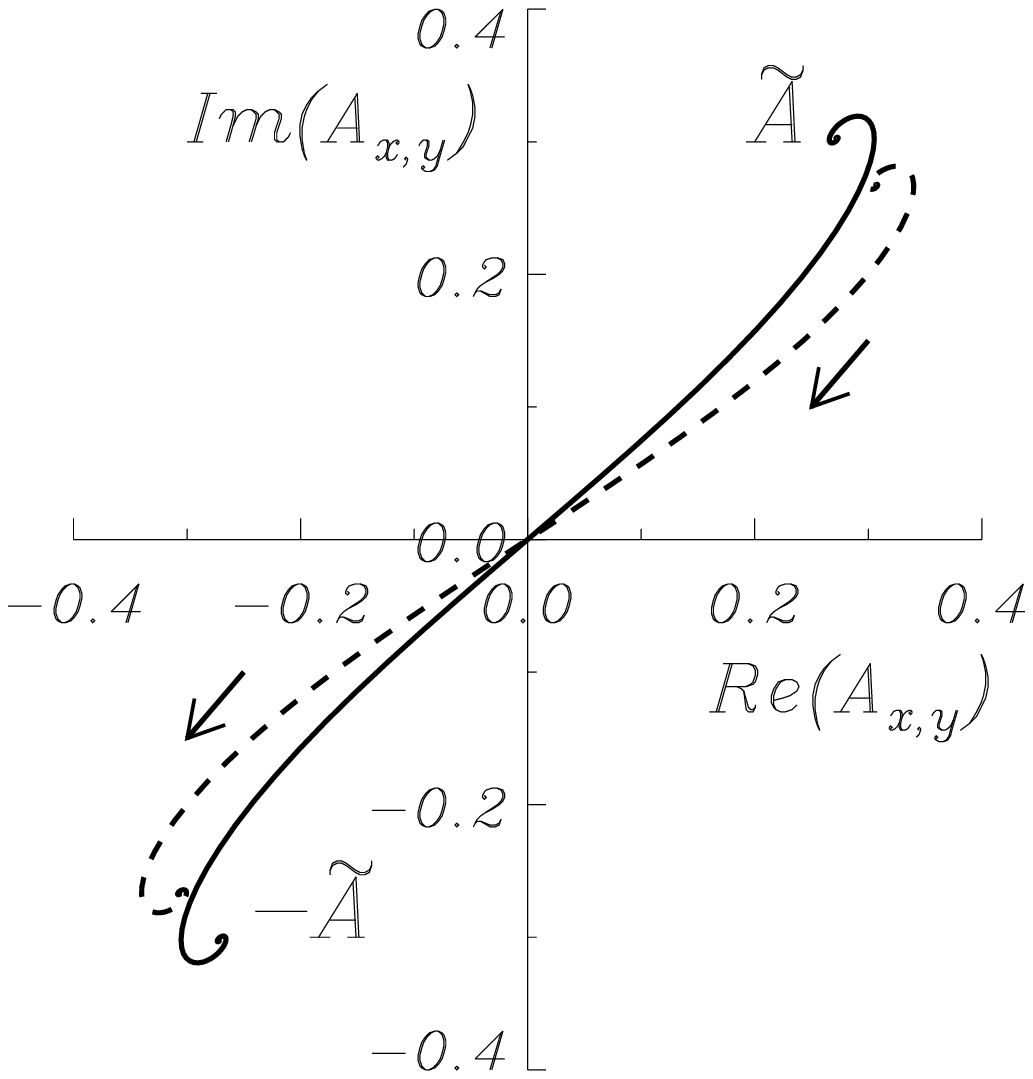,width=4.5in}}
\centerline{Figure 3b}

\newpage

\centerline{\psfig{figure=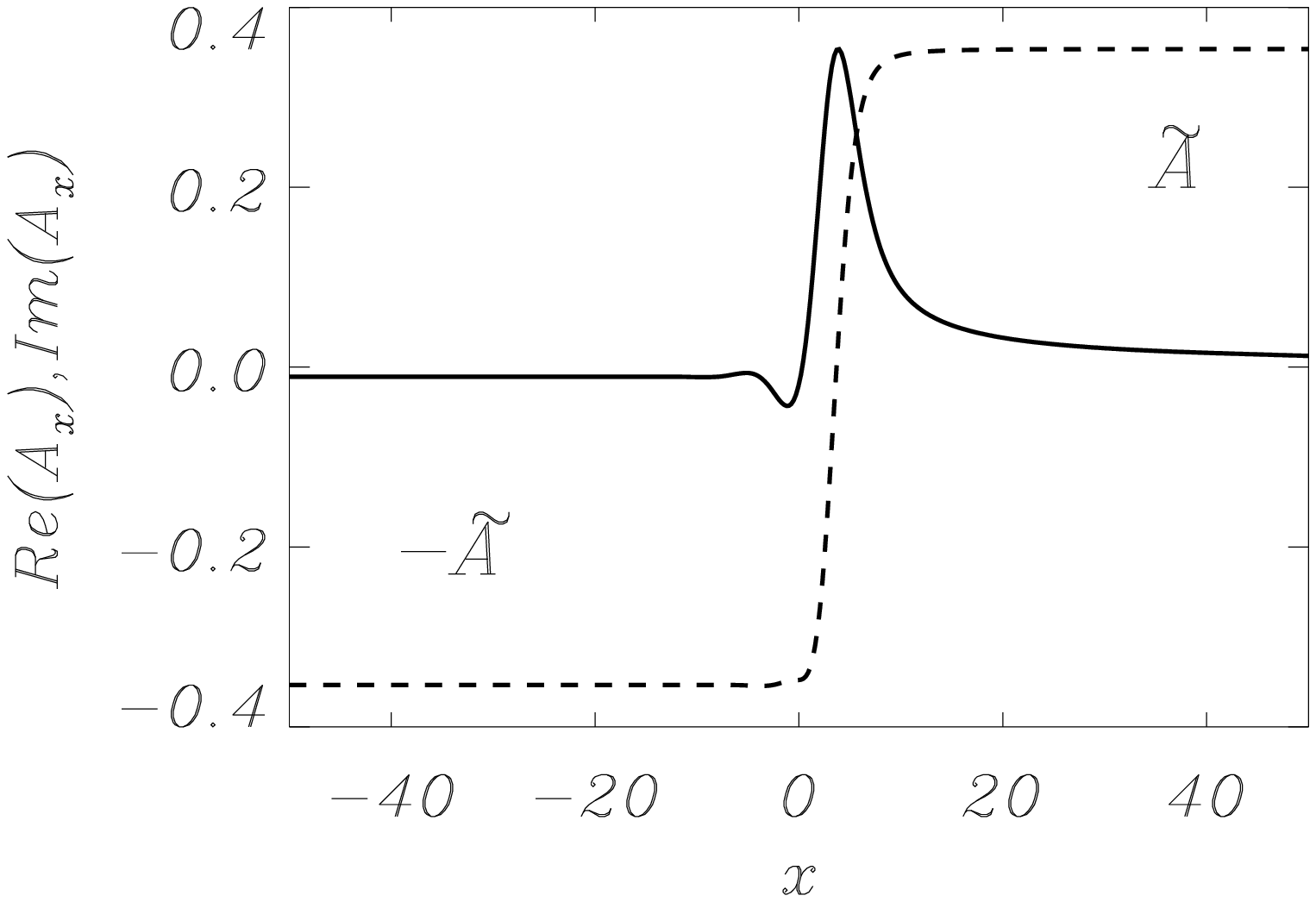,width=5in}}
\centerline{Figure 4a}
\centerline{\psfig{figure=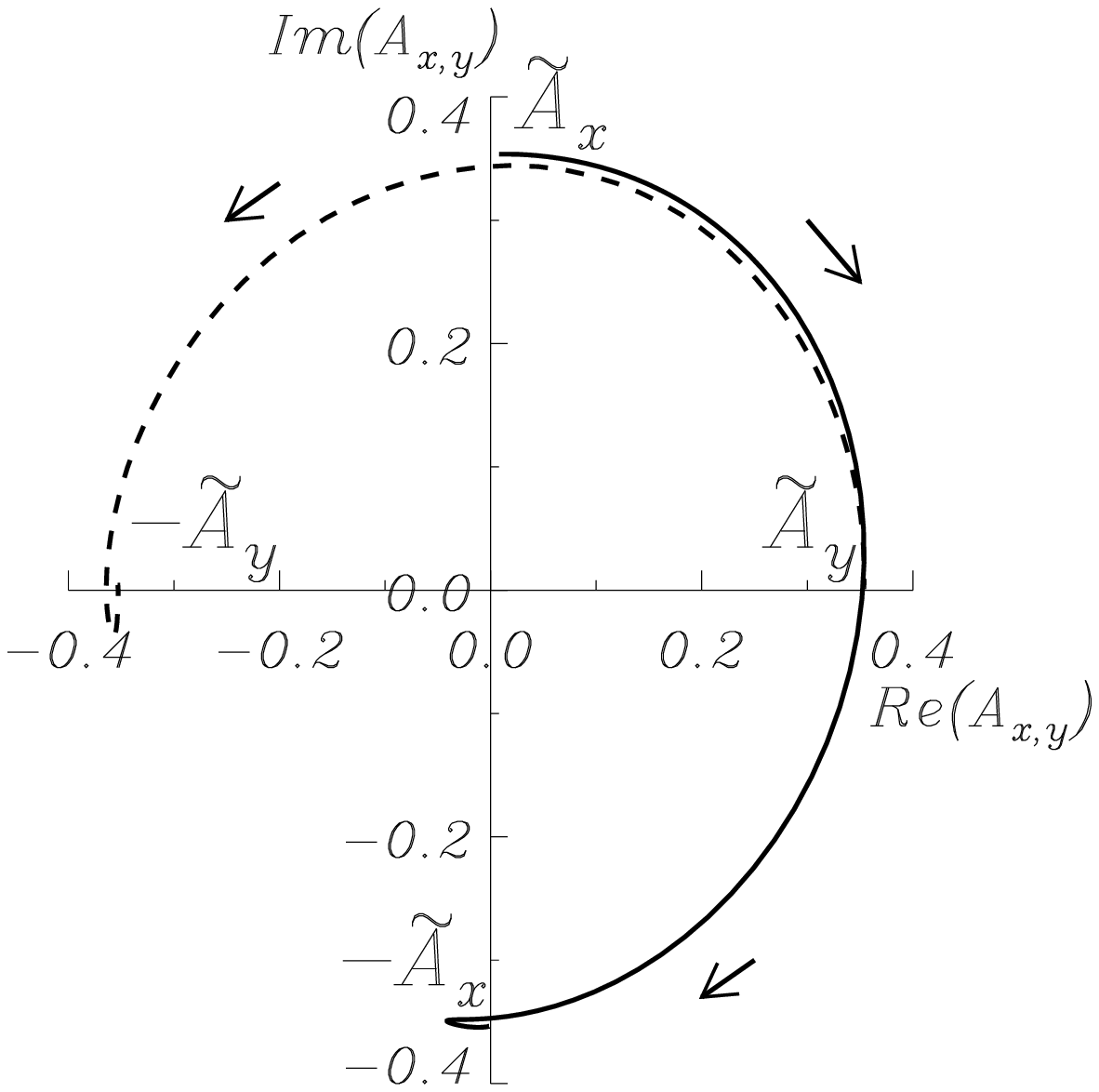,width=4.5in}}
\centerline{Figure 4b}

\newpage

\centerline{\psfig{figure=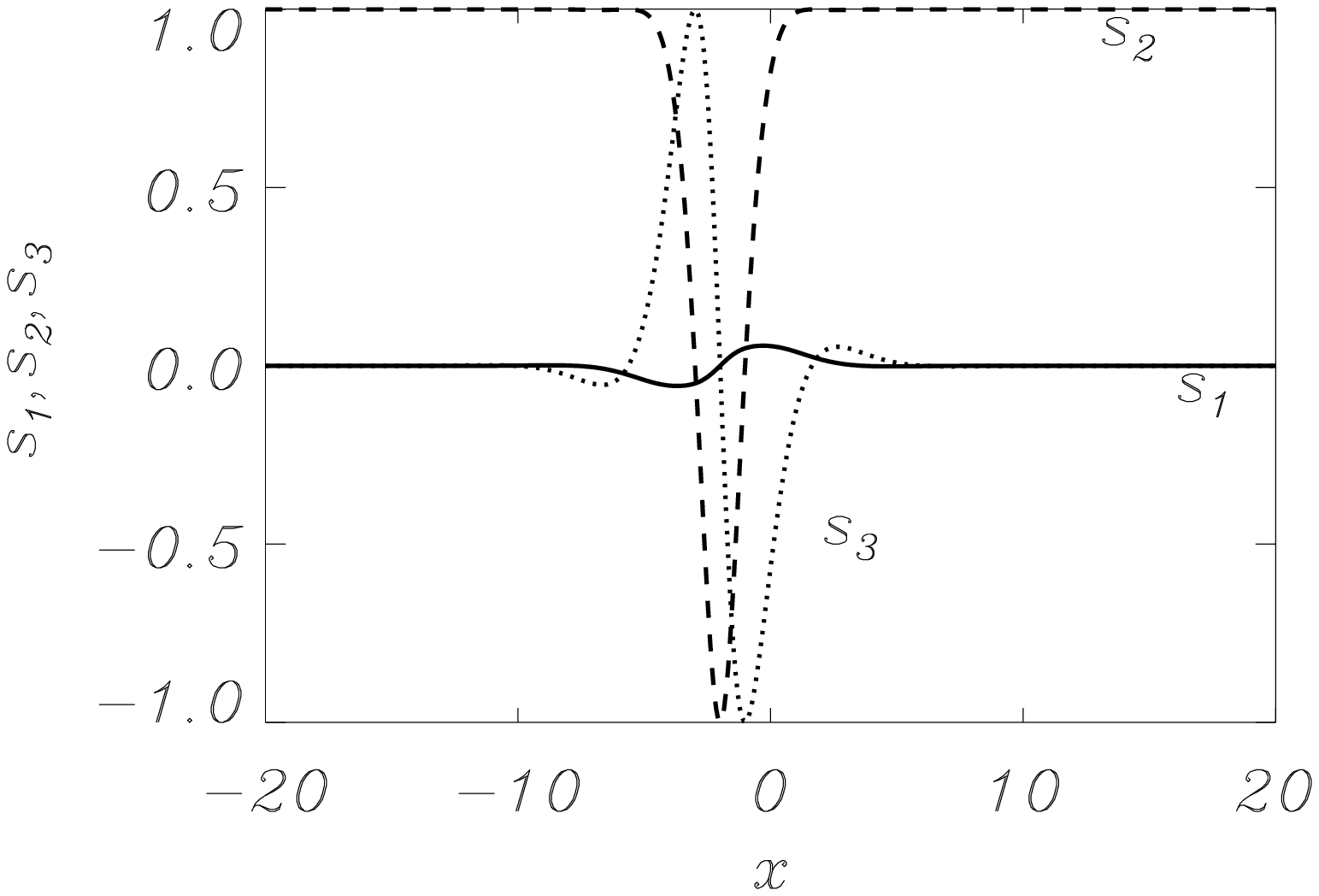,width=5in}}
\centerline{Figure 5a}
\centerline{\psfig{figure=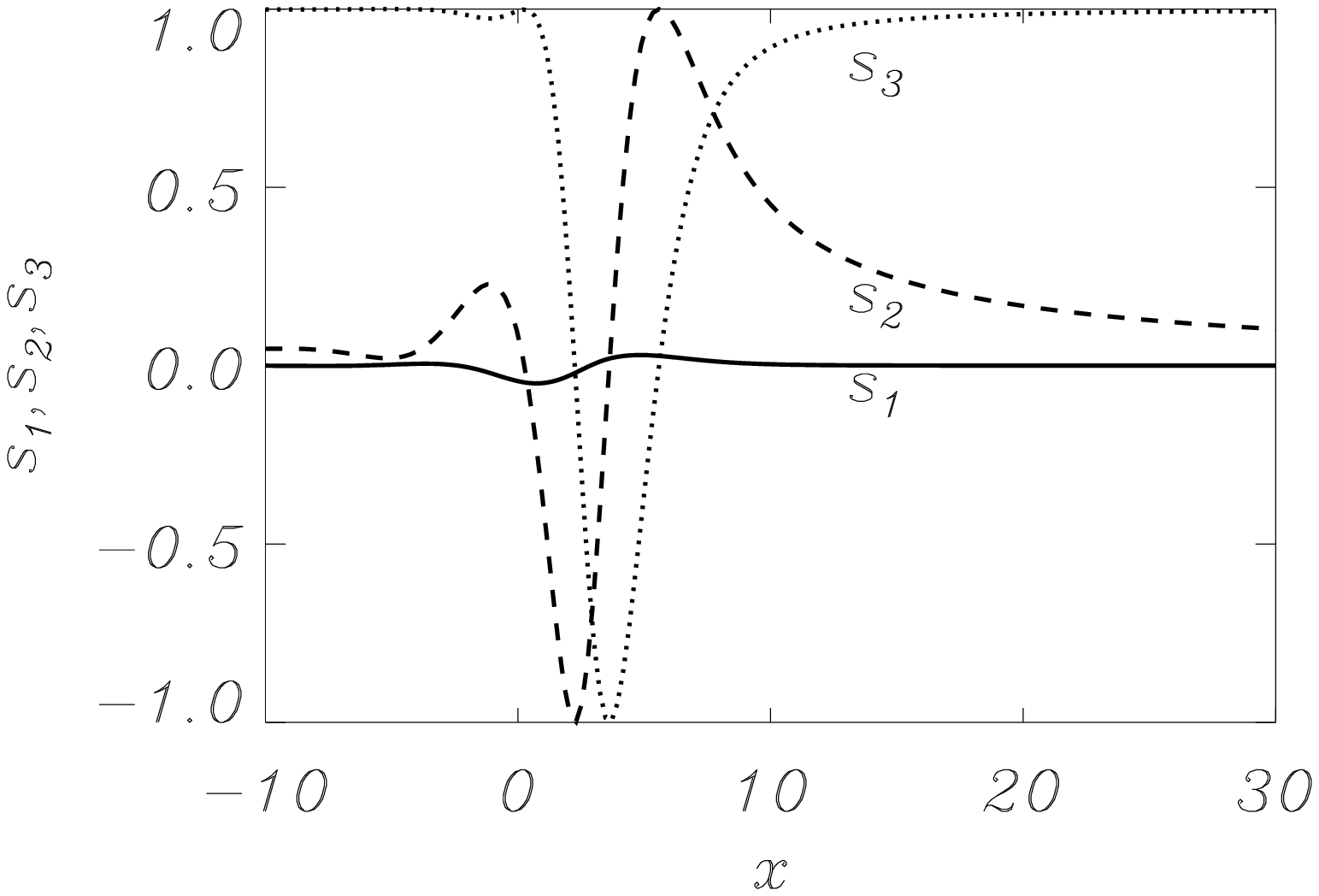,width=5in}}
\centerline{Figure 5b}

\newpage

\centerline{\psfig{figure=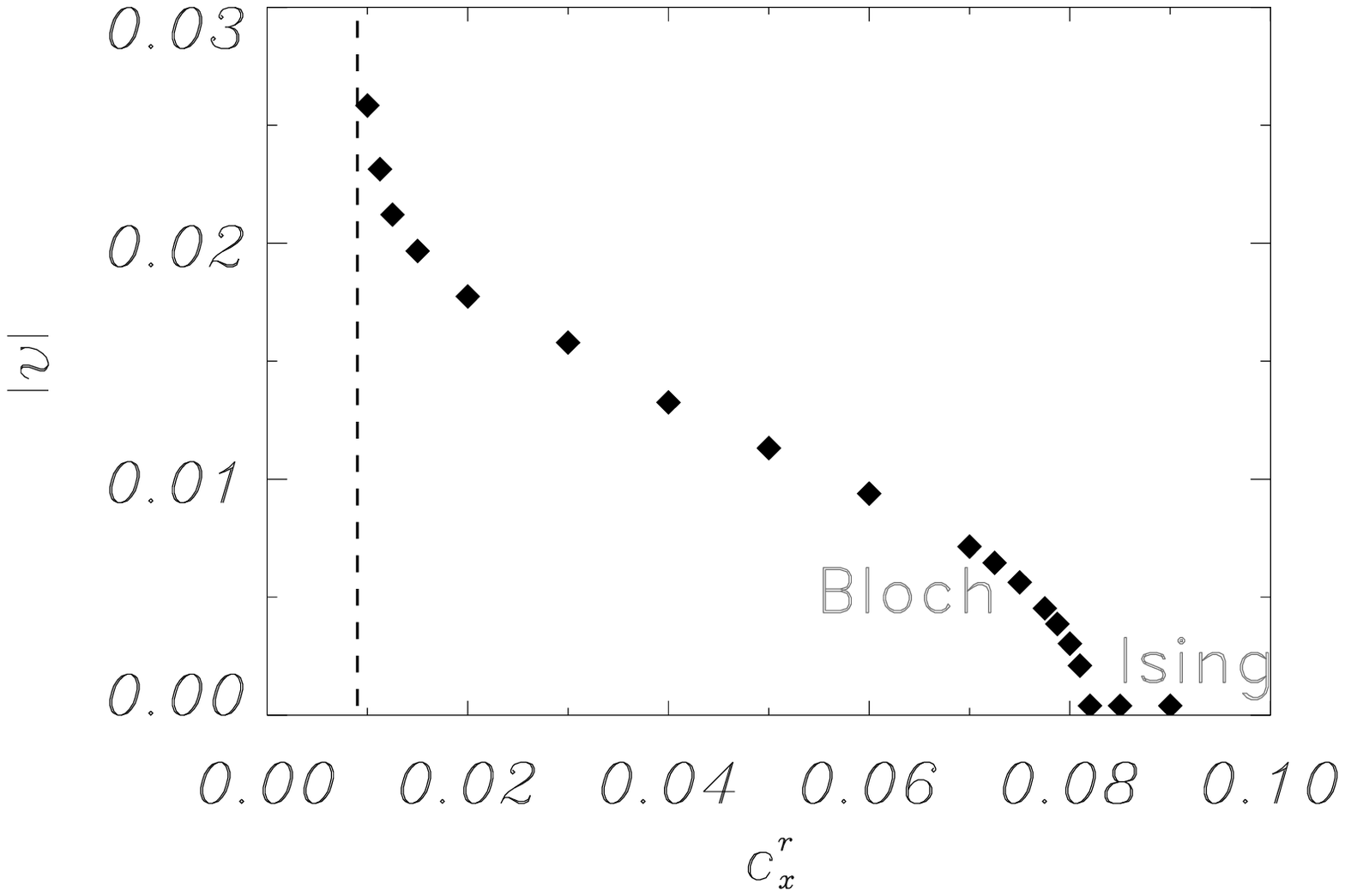,width=5in}}
\centerline{Figure 6}
\centerline{\psfig{figure=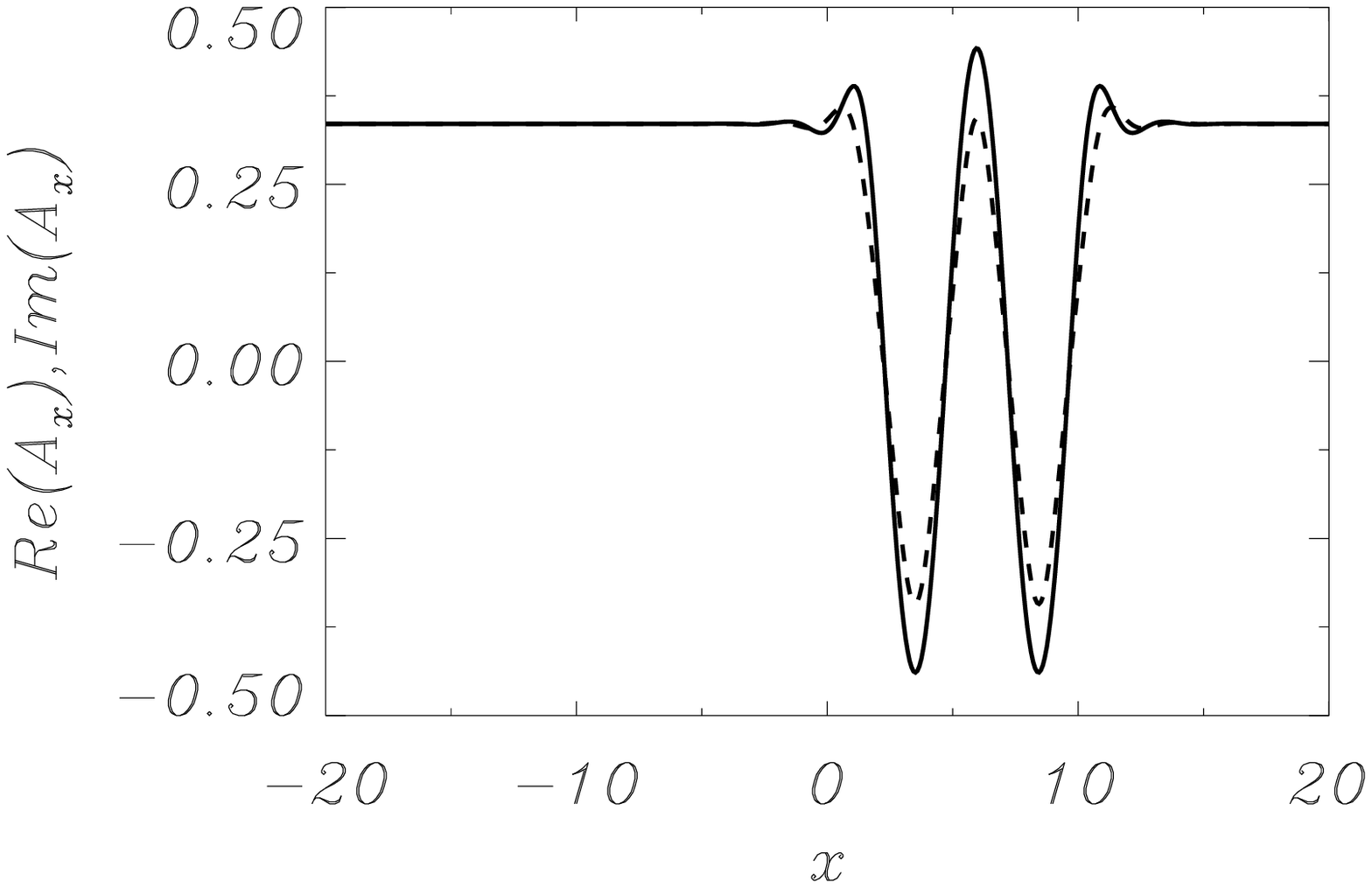,width=5in}}
\centerline{Figure 7a}

\newpage

\centerline{\psfig{figure=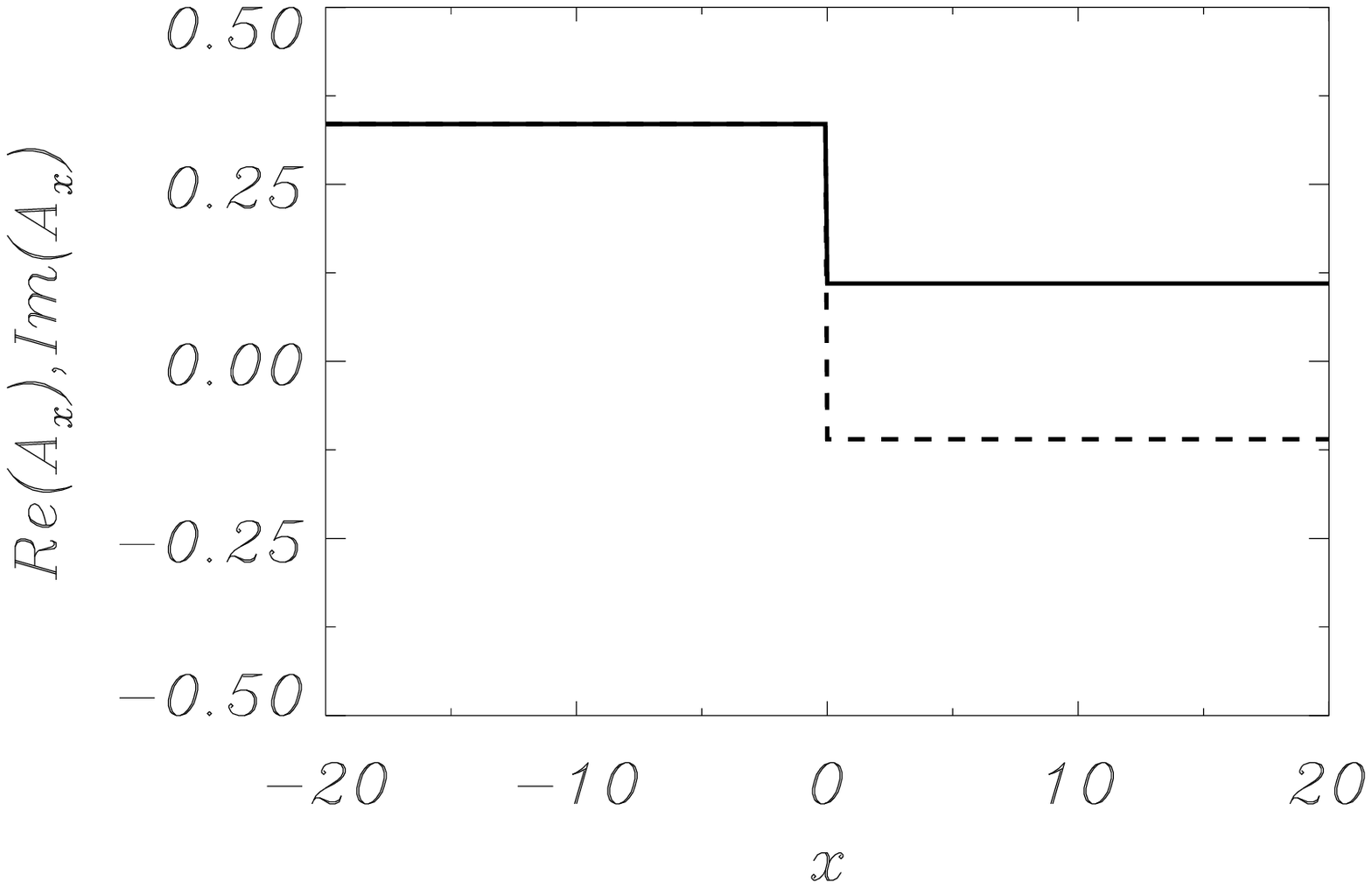,width=5in}}
\centerline{Figure 7b}
\centerline{\psfig{figure=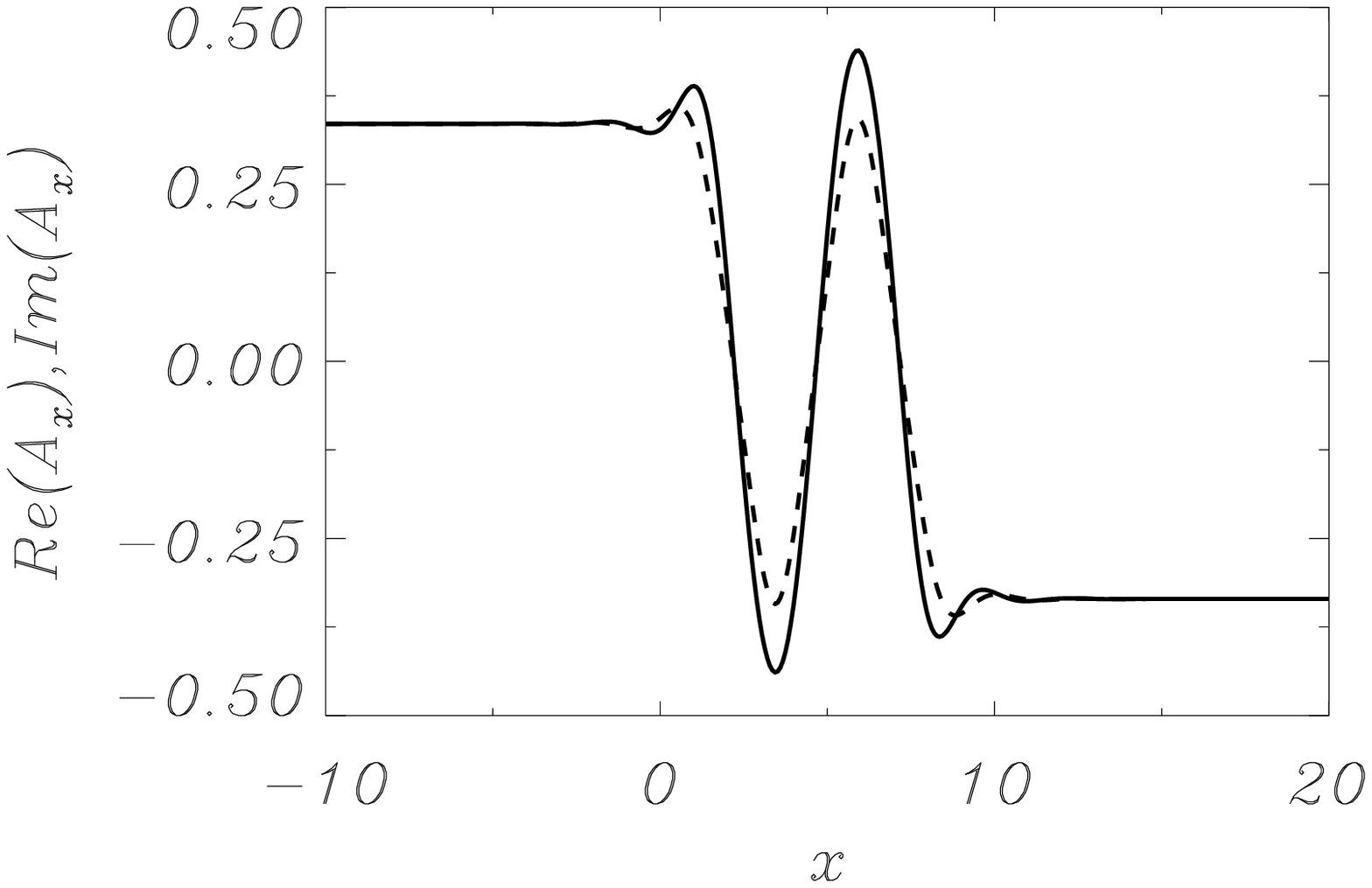,width=5in}}
\centerline{Figure 7c}

\newpage

\centerline{\psfig{figure=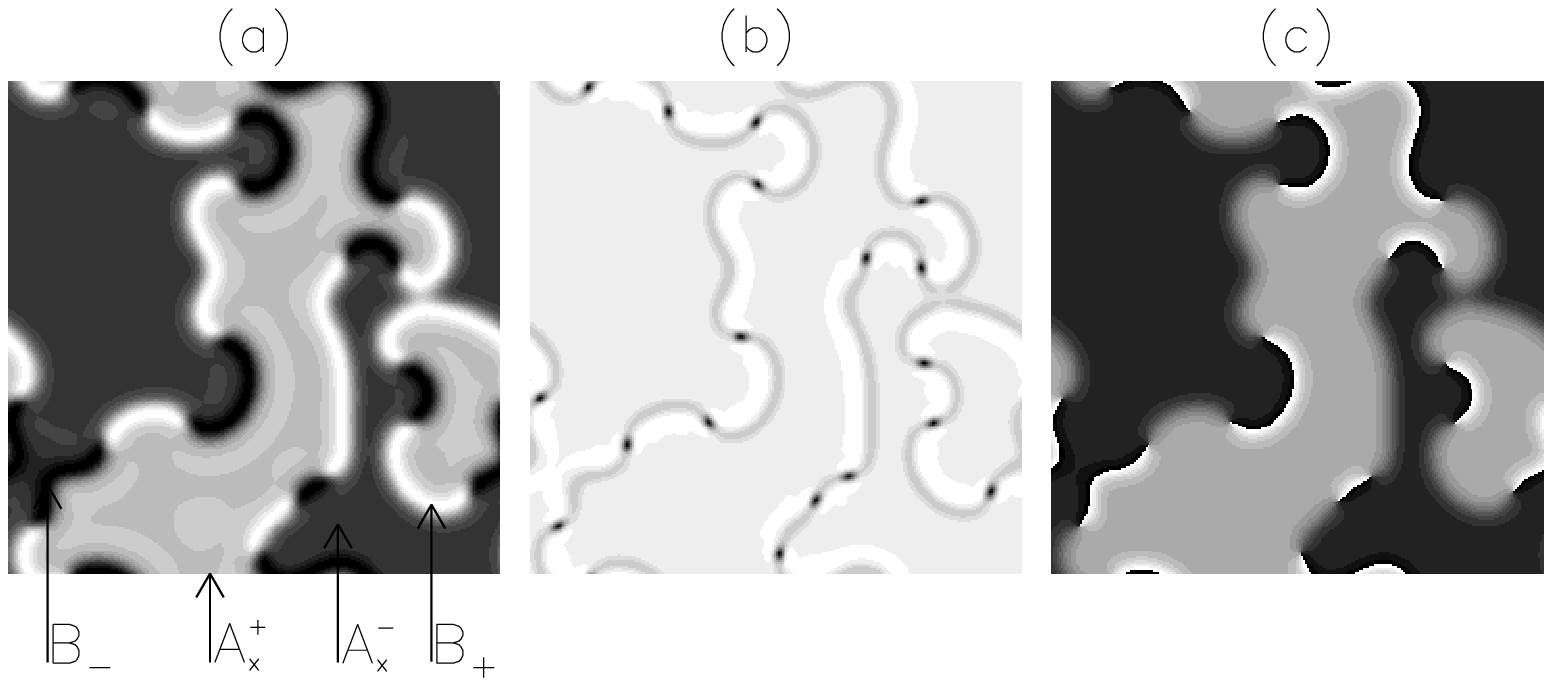,width=5in}}
\centerline{Figure 8}

\centerline{\psfig{figure=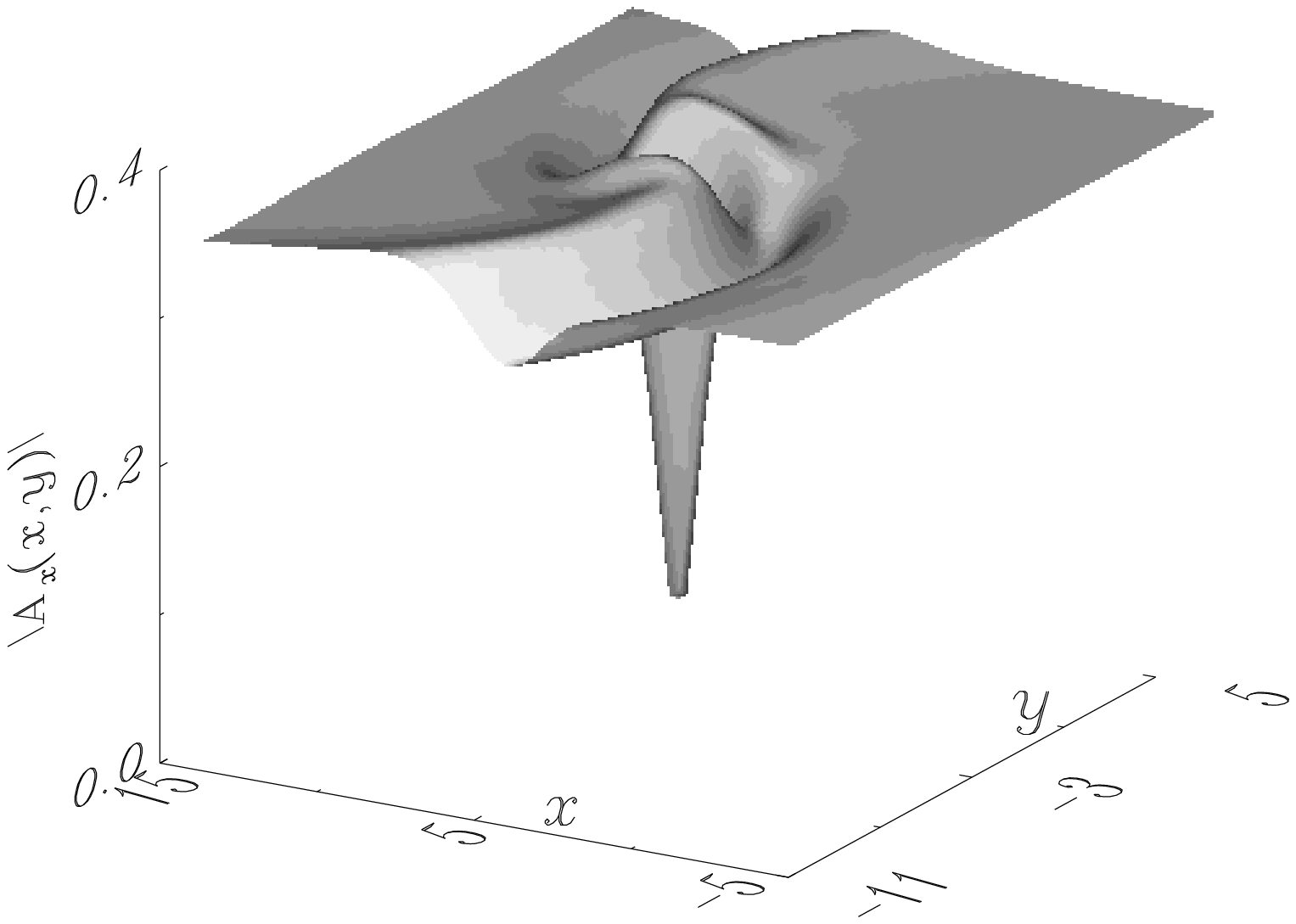,width=5in}}
\centerline{Figure 9}

\newpage

\centerline{\psfig{figure=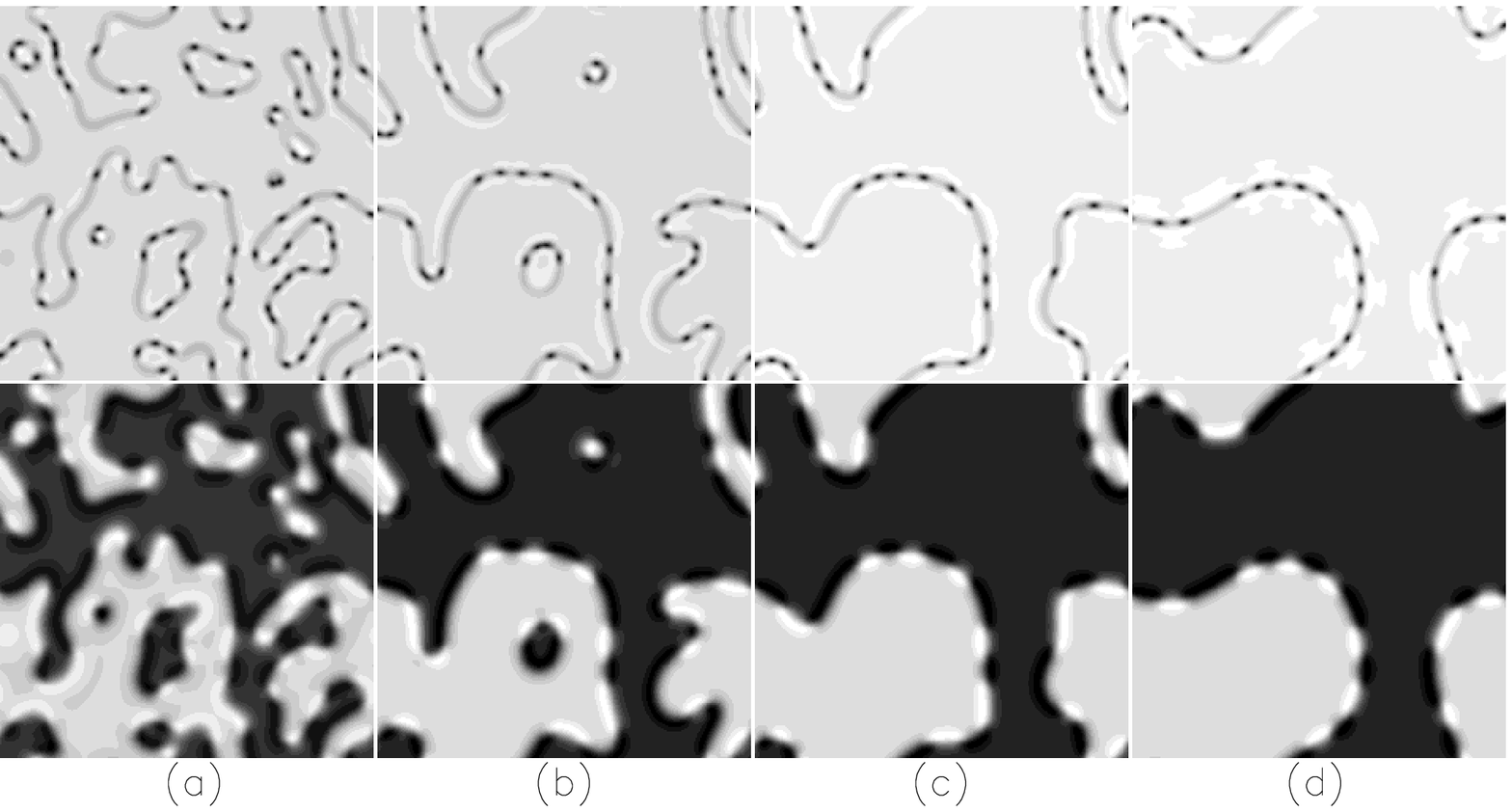,width=5in}}
\centerline{Figure 10}
\centerline{\psfig{figure=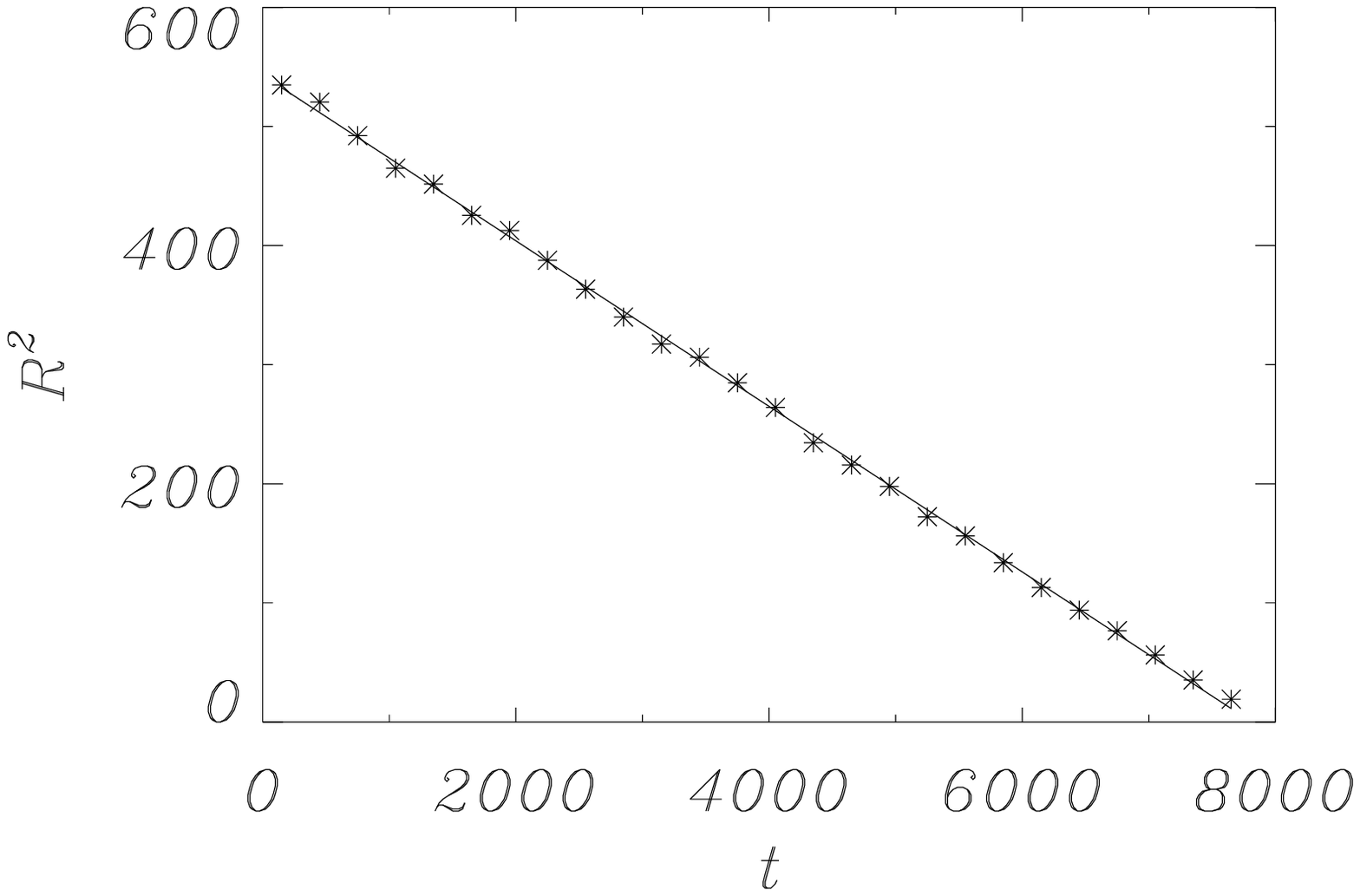,width=5in}}
\centerline{Figure 11}

\newpage

\centerline{\psfig{figure=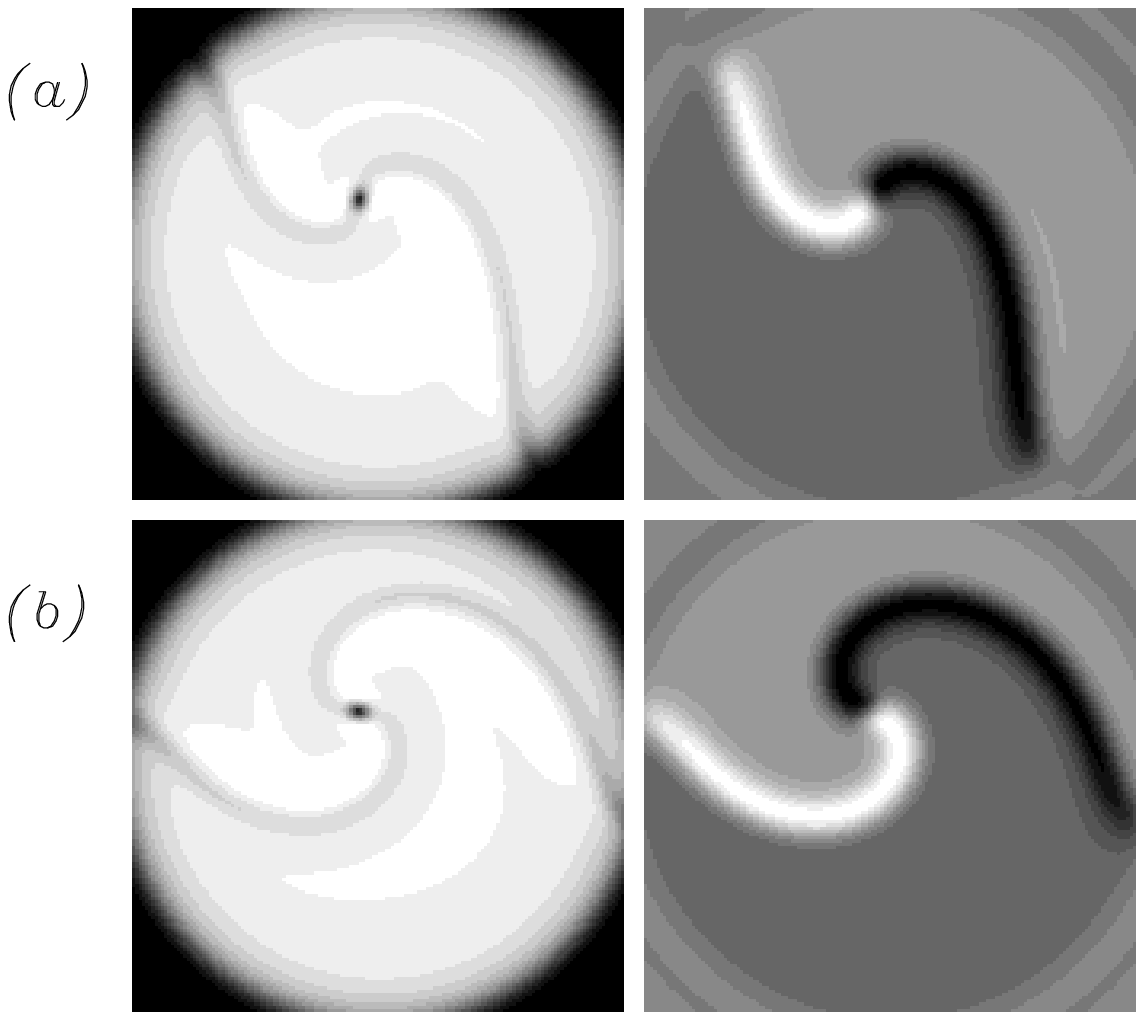,width=5in}}
\centerline{Figure 12}

\newpage

\centerline{Figure 13}
\centerline{\psfig{figure=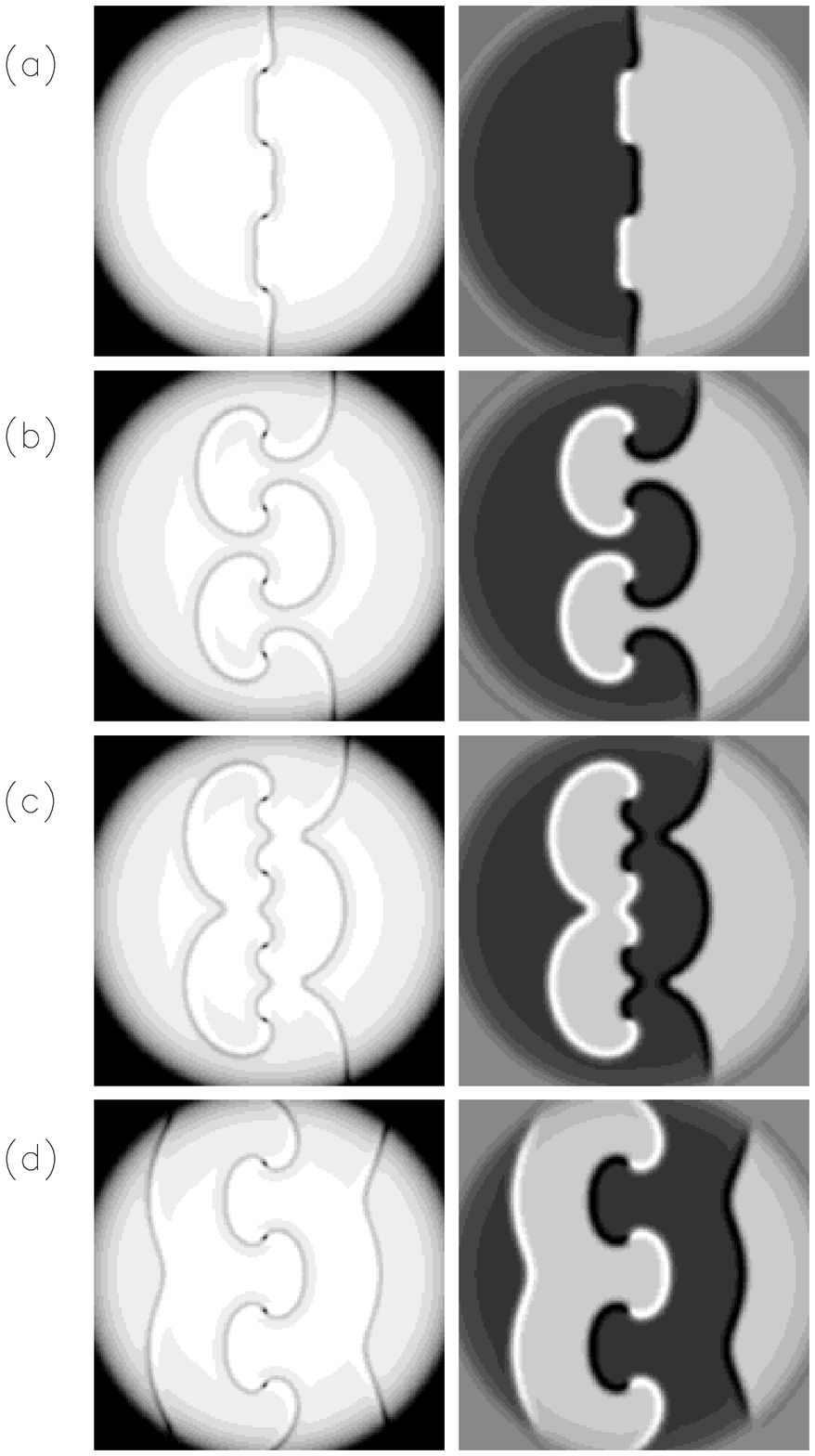,width=5in}}

\newpage

\centerline{\psfig{figure=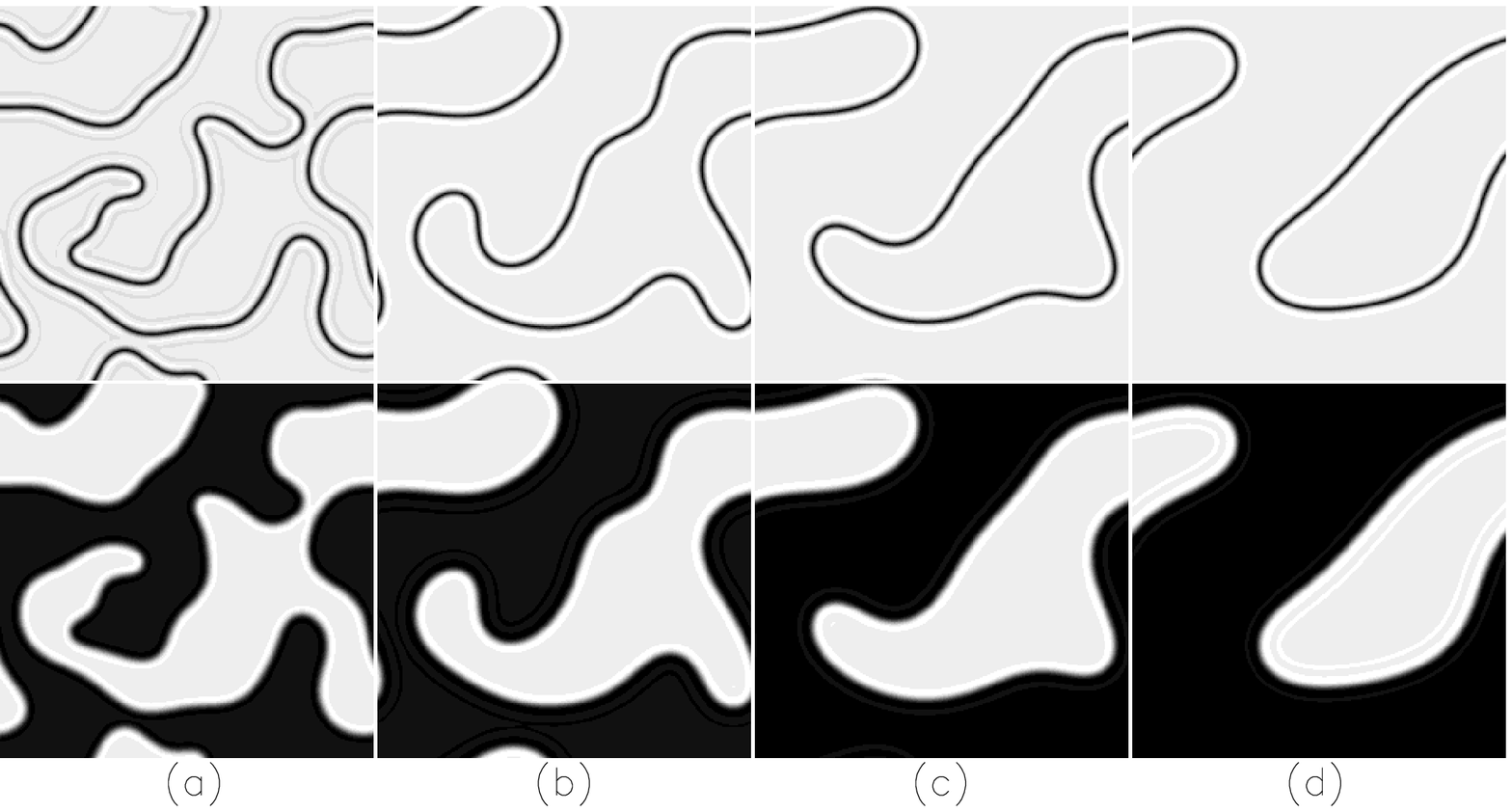,width=5in}}
\centerline{Figure 14}
\centerline{\psfig{figure=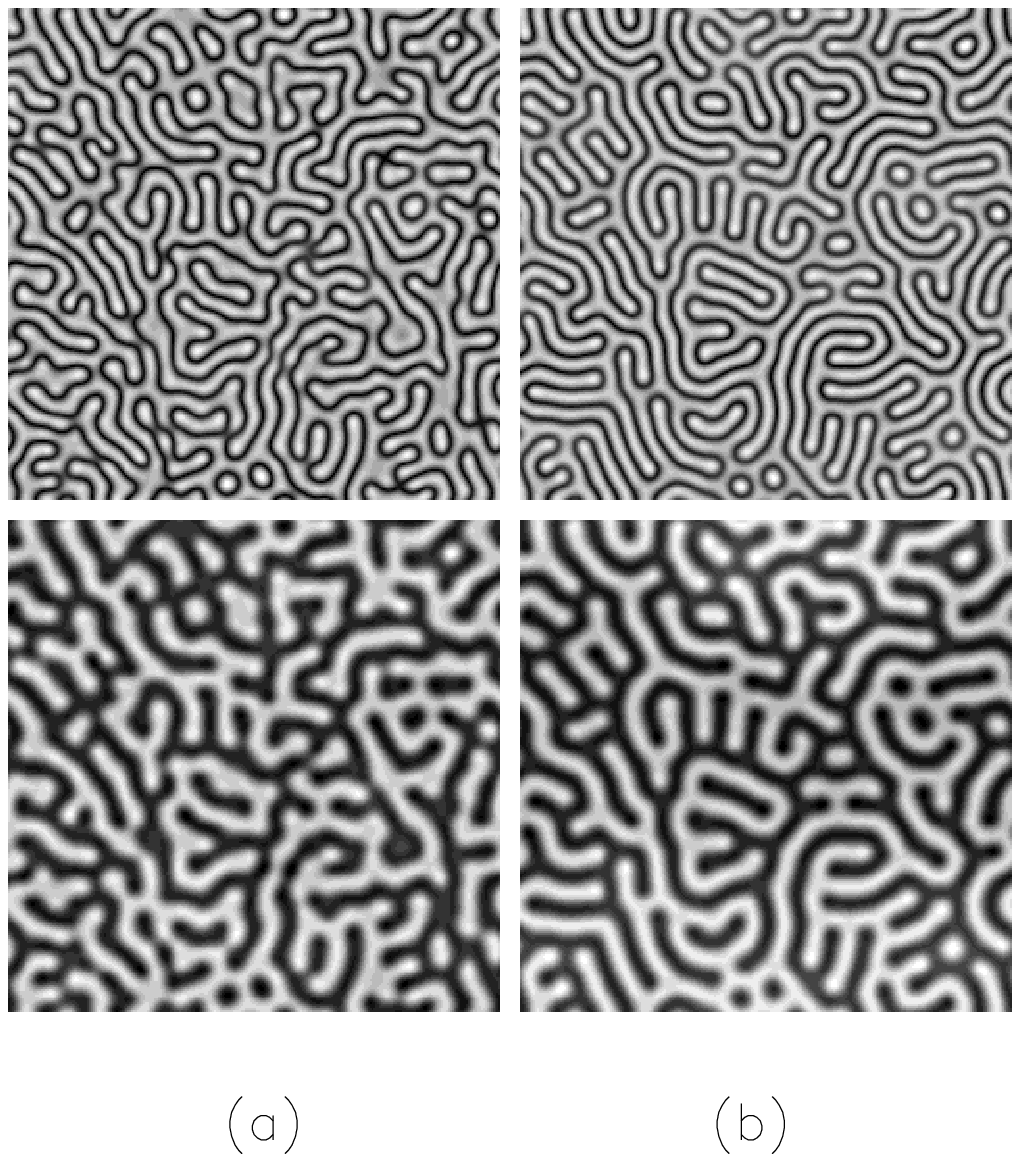,width=4.5in}}
\centerline{Figure 15}
\centerline{\psfig{figure=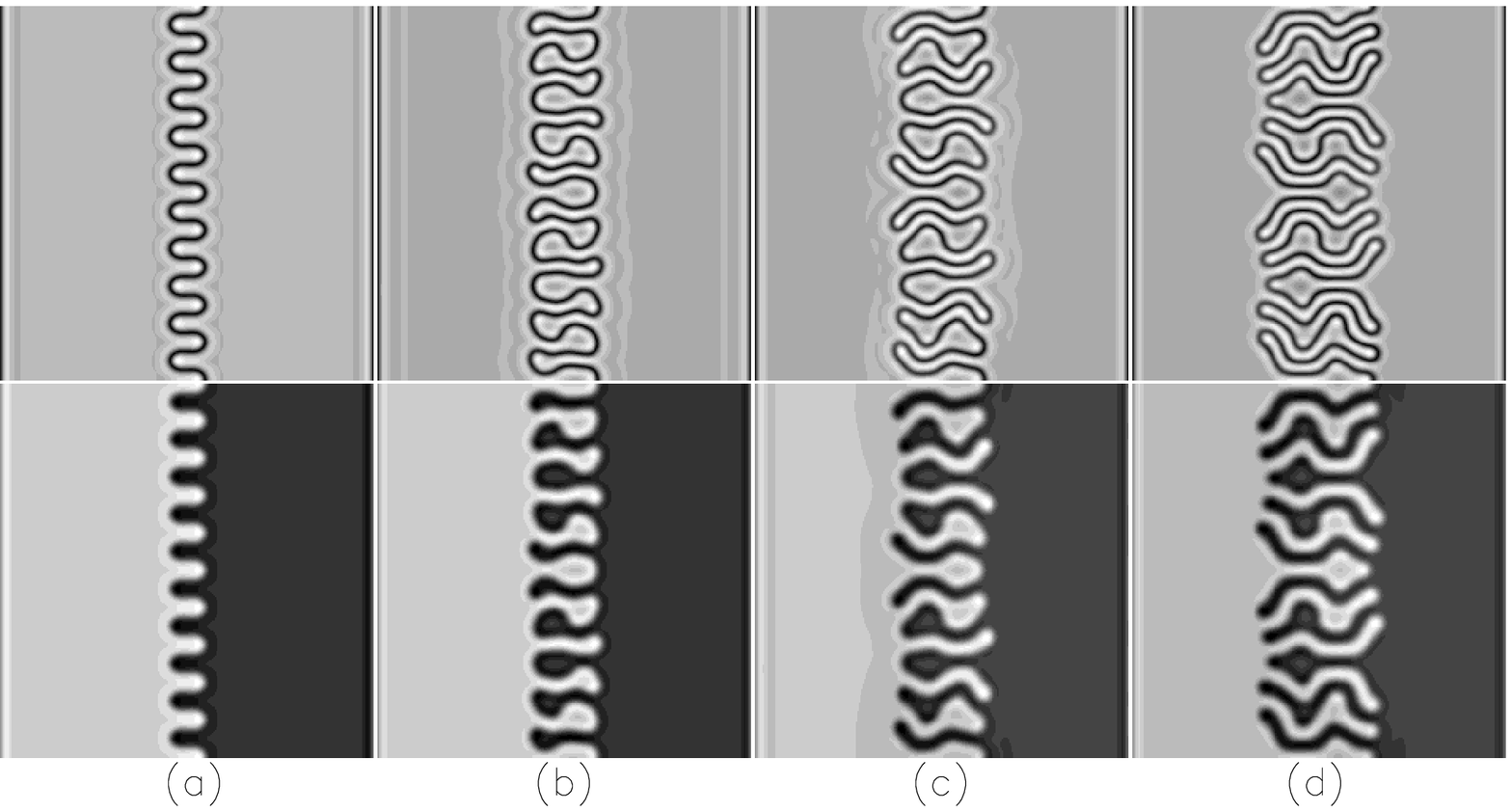,width=5in}}
\centerline{Figure 16}


\begin{thebibliography}{99}


\bibitem[\dagger]{gonzalo}  Permanent address: Departamento de 
F\'{\i}sica, Facultad de Ciencias Exactas y Naturales, Universidad 
Nacional de Mar del Plata. Funes 3350 (7600) Mar del Plata, Argentina. (izus@mdp.edu.ar). Member of CONICET  (Argentine).

\bibitem[*]{www} http://www.imedea.uib.es/PhysDept.

\bibitem{JOSAB99} J. Opt. Soc. Am. B {\bf 16}, (1999, special issue).

\bibitem{quantum} M. D. Reid, and P. Drummond, Phys. Rev. Lett. {\bf 60},
2731 (1988); J. Mertz, T. Debuisschert, A. Heidmann, C. Fabre and E. Giacobino,
Opt. Lett. {\bf 16}, 1234 (1991); P. G. Kwiat, K. Mattle, H. Weinfurter, A. Zeilinger,
A. V. Sergienko and Y. Shih, Phys. Rev. Lett. {\bf 75}, 4337 (1995).

\bibitem{Lugiatoreview} L. Lugiato, A. Gatti, and H. Wiedemann,
{\it Quantum Fluctuations and Nonlinear Optical Patterns} edited
by S. Reynaud, E. Giacobino, and J. Zinn-Justin, Les Houches, Session LXIII, 1995
(1997 Elseiver Science B. V.).

\bibitem{pinos} Proceedings of the Euroconference: Patterns in nonlinear
optical systems, Alicante (Spain) 1998; see also the special issue of
{\em Journal of Optics B: Quantum Semiclass. Opt.}, {\bf 1}, (1999).

\bibitem{oppo94} G-L Oppo, M. Brambilla, and L. A. Lugiato, Phys. Rev. A {\bf
49}, 2028 (1994).

\bibitem{valcarcel96} G. C. de Valcarcel, K. Staliunas, E. Roldan, and V. J.
Sanchez-Morcillo, Phys. Rev. A {\bf 54}, 1609 (1996).

\bibitem{fabr99} M. Vaupel, A. Maître, and C. Fabre, Phys. Rev. Lett. {\bf 83},
5278 (1999).

\bibitem{oppo01} G-L Oppo, A. J. Scroggie, and W. J. Firth, to appear in 
Phys. Rev. E (2001)

\bibitem{Berre} M. Le Berre, D. Leduc, E. Ressayre, and A. Tallet,
J. Opt. B: {\it Quantum and Semiclass. Opt.} {\bf 1}, 153 (1999);
M. Tlidi, M. Le Berre, A. Ressayre, A. Tallet, and L. Di Menza,
Phys. rev. A {\bf 61}, 43806 (2000). 

\bibitem{op2} G. Iz\'us, M. Santagiustina, and M. San Miguel,
Opt. Lett. {\bf 25}, 1454 (2000).

\bibitem{tril97} S. Trillo, M. Haelterman, and A. Sheppard, Opt. Lett. {\bf 22},
970 (1997).

\bibitem{long97} S. Longhi, Phys. Scr. {\bf 56}, 611 (1997).

\bibitem{op1} M. Santagiustina, P. Colet, M. San Miguel, and D. Walgraef,
Opt. Lett. {\bf 23}, 1167 (1998).

\bibitem{stal98} K. Staliunas and V. S\'anchez-Morcillo, Phys. Rev. A {\bf 57},
1454 (1998).

\bibitem{oppo99} G.L. Oppo, A.J. Scroggie, and W.J. Firth, {\em Journal of
Optics B: Quantum Semiclass. Opt.} {\bf 1}, 133 (1999);

\bibitem{Mason98} E. J. Mason and N.C. Wong, Opt. Lett. {\bf 23}, 1733 (1998).

\bibitem{Fabre00} C. Fabre, E. Mason, and N. Wong,  Opt. Comm. {\bf 170}, 
299 (1999).

\bibitem{Nos99} G. Iz\'us, M. Santagiustina, M. San Miguel, and P. Colet,
J. Opt. Soc. Am. B {\bf 16}, 1592 (1999).

\bibitem{kutz99} N. Kutz, T. Ernaux, S. Trillo, and M. Haelterman, Journ. Opt.
Soc. Am  B {\bf 16}, 1936 (1999).

\bibitem{oppo98} G-L. Oppo, A.J. Scroggie, and W.J. Firth, {\em European
Quantum Electronics Conference Digest}, (Glasgow, UK) 245 (1998).

\bibitem{Coullet90} P. Coullet, J. Lega, B. Houchmanzadeh, and J. Lajzerowicz,
Phys. Rev. Lett. {\bf 56}, 1352 (1990).

\bibitem{thorsten} E. G. Westhoff, V. Kneisel, Y. A. Logvin, T. Ackermann, and
W. Lange, J. Opt. B: {\it Quantum and Semiclass. Opt.} {\bf 2}, 386 (2000)


\bibitem{lede98} U. Peschel, D. Michaelis, C. Etrich, and F. Lederer, Phys.
Rev. E {\bf 58}, R2745 (1998).

\bibitem{mand98} M. Tlidi, P. Mandel, and R. Lefever, Phys. Rev. Lett. {\bf 81},
979 (1998).

\bibitem{stal98b} K. Staliunas and  V. S\'anchez-Morcillo, Phys. Lett. A {\bf 241},
28 (1998).

\bibitem{rafa} R. Gallego, M. San Miguel and R. Toral, Phys. Rev. E {\bf 61},
2241 (2000).

\bibitem{tara98} V. Taranenko, K. Staliunas, and C. Weiss, Phys. Rev. Lett.
{\bf 81}, 2236 (1998).

\bibitem{roza96} N. N. Rozanov, Progress in Optics, E. Wolf Ed. {\bf 35}, 1
(1996).

\bibitem{bloch} D. Michaelis, U. Peschel, F. Lederer, D. V. Skryabin,
and W. Firth, to appear in Phys. Rev. Lett. (2001); G. de Valcarcel, and
K. Staliunas, (submitted).

\bibitem{kimb90} H. J. Kimble, Quantum fluctuations in quantum
optics-squeezing and related phenomena, (J. Dalibard, J. M. Raimond, J.
Zinn-Justin eds.).  Elsevier Science Publishers, (1992).

\bibitem{Lee90} D. Lee and N. Wong, Appl. Phys. B {\bf 66}, 133 (1998).

\bibitem{phot} B. E. A. Saleh, M. C. Teich, "The fundamentals of photonics", Wiley,
Chap. 6 (1991).

\bibitem{falk71} J. Falk, IEEE Jour. Quant. El. {\bf QE-7}, 230 (1971).

\bibitem{byer91} R. Eckardt, C. D. Nabors, W.J. Kozlowsky, and R. L. Byer,
Journ. Opt. Soc. Am  B {\bf 8}, 646 (1991).

\bibitem{fabr93} T. Debuisschert, A. Sizmann, E. Giacobino, and C. Fabre,
Journ. Opt. Soc. Am  B {\bf 10}, 1668 (1993).

\bibitem{opos} J-Y Zhang, J. Y. Huang, Y. R. Shen, Laser Science and Technology,
{\bf 19}, (Harwood Ac.) (1995); C.L. Tang, L. K. Cheng, Laser Science and
Technology,  {\bf 20} (Harwood Ac.) (1995).

\bibitem{numerical} Eqs. (\ref{master}) have been integrated using the 
algorithm described in  ref. \cite{pre}. In 1D we take a grid of 2048 
samples, with periodic boundary conditions. For all cases the grid space 
was $\Delta x=0.078125$ and the integration step was $\Delta t=0.001$.
For the 2D case  we use a grid of 256x256 samples with grid space
$\Delta x=\Delta y=0.3125$ and time step $\Delta t=0.02$. In some
cases we use, instead of periodic boundary conditions, a flat-top
pump beam $E_0(x,y)$ (details of such pump beam are given in ref.
\cite{pre}).

\bibitem{Tutu97} H. Tutu, Phys. Rev. E {\bf 56}, 5036 (1997).

\bibitem{frisc94} T. Frisch, S. Rica, P. Coullet, and J. M. Gilli, Phys. 
Rev. Lett. {\bf 72}, 1471 (1994).

\bibitem{stal99}C. O. Weiss, M. Vaupel, K. Staliunas, G. Slekys, 
and V. B. Taranenko, Appl. Phys. B, {\bf 68}, 151 (1999).

\bibitem{lugi88} L. A. Lugiato, C. Oldano, Phys. Rev. A, {\bf 37},
3896 (1988).

\bibitem{pre} M. Santagiustina, P. Colet, M. San Miguel, D. Walgraef,
Phys. Rev. E {\bf 58}, 3843 (1998);

\end{thebibliography}
\end{document}